\documentclass[12pt, a4paper]{article}
\usepackage{jheppub_mod}
\usepackage{bm} 
\usepackage{subcaption}
\usepackage{graphicx} 
\usepackage{comment}
\usepackage{physics}
\usepackage{multirow}
\usepackage[usenames,dvipsnames,svgnames,table]{xcolor}
\definecolor{hblue}{rgb}{0,0,0.575}
\allowdisplaybreaks
 \title{Sign Flip Triangulations of the Amplituhedron}
 \author[a,b]{Ryota Kojima,}
\author[c]{Cameron Langer,}
\affiliation[a]{Department of Particle and Nuclear Physics, SOKENDAI (The Graduate University for Advanced Studies)\\ Tsukuba, Ibaraki, 305-0801, Japan}
\affiliation[b]{KEK Theory Center, Tsukuba, Ibaraki, 305-0801, Japan}
\affiliation[c]{Center for Quantum Mathematics and Physics (QMAP), University of California, Davis, CA, USA}
\emailAdd{cklanger@ucdavis.edu}
\emailAdd{ryota@post.kek.jp}
\abstract{
We present new triangulations of the $m=4$ amplituhedron relevant for scattering amplitudes in planar $\mathcal{N}=4$ super-Yang-Mills, obtained directly from the combinatorial definition of the geometry. Using the ``sign flip'' characterization of the amplituhedron, we reproduce the canonical forms for the all-multiplicity next-to-maximally helicity violating (NMHV) and next-to-next-to-maximally helicity violating ($\text{N}^2$MHV) tree-level as well as the NMHV one-loop cases, without using any input from traditional amplitudes methods. Our results provide strong evidence for the equivalence of the original definition of the amplituhedron \cite{Arkani-Hamed:2013jha} and the topological one \cite{Arkani-Hamed:2017vfh}, and suggest a new path forward for computing higher loop amplitudes geometrically. In particular, we realize the NMHV one-loop amplituhedron as the intersection of two amplituhedra of lower dimensionality, which is reflected in the novel structure of the corresponding canonical form.
}
\begin{document}
\begin{flushright}
KEK-TH-2177
\end{flushright}
\maketitle
\newpage

\section{Introduction}
Scattering amplitudes have been a continuous source of insight into the hidden structure and simplicity underlying perturbative quantum field theory. Recent years have revealed an unexpected and surprising connection between the $S$-matrix in an increasingly wide variety of theories and a broad notion of ``positive geometry'' \cite{Arkani-Hamed:2013jha, Arkani-Hamed:2013kca, Arkani-Hamed:2017vfh,Arkani-Hamed:2017tmz, Arkani-Hamed:2017fdk, Arkani-Hamed:2017mur, Arkani-Hamed:2018ign,Arkani-Hamed:2019mrd,Arkani-Hamed:2019vag,Arkani-Hamed:2019rds}. The complete geometric reformulation of the Feynman diagram approach to calculating amplitudes was accomplished in $\mathcal{N}=4$ supersymmetric-Yang-Mills (sYM) in the planar limit with the definition of the amplituhedron. This remarkable generalization of polytopes and the positive Grassmannian is conjectured to contain all of the complexities of tree-level amplitudes and loop-level integrands of the theory by associating to the geometry a canonical differential form, defined by having logarithmic singularities on all boundaries (of all co-dimensionality). More broadly, the key idea that scattering amplitudes can be understood as differential forms on kinematical space extends to more general theories \cite{Arkani-Hamed:2014via,He:2018okq,He:2018pue,He:2018svj,Banerjee:2018tun,Salvatori:2018fjp,Salvatori:2019phs,Herderschee:2019wtl,Aneesh:2019cvt}. The amplituhedron has been utilized to probe all-loop order information about the loop integrand inaccessible from any diagrammatic approach \cite{Arkani-Hamed:2018rsk, Langer:2019iuo}, symbol alphabets and branch cut structure \cite{Prlina:2017tvx,Prlina:2017azl} and has been explored from a variety of physical and mathematical perspectives \cite{Arkani-Hamed:2014dca, Franco:2014csa, Bai:2014cna, Lam:2014jda,Lam:2015uma,Karp:2015duv,Ferro:2015grk, Ferro:2016ptt,Dennen:2016mdk, An:2017tbf, Rao:2017fqc, Galashin:2017onl,Kojima:2018qzz,Salvatori:2018aha,Galashin:2018fri, Damgaard:2019ztj,YelleshpurSrikant:2019meu,Rao:2019wyi,Lukowski:2019sxw,Lukowski:2019kqi}.

Calculating amplitudes or loop integrands starting from the amplituhedron requires the construction of the canonical form associated to the geometry. To date there is one completely general and in principle straightforward way to do this: by triangulating the amplituhedron into elementary cells for which the canonical form is easy to compute, and subsequently summing the individual pieces \cite{Arkani-Hamed:2013kca}. Although triangulation will be the primary approach used in this paper, some interesting alternative methods for computing the canonical form have been proposed \cite{Arkani-Hamed:2014dca, Enciso:2014cta, Ferro:2018vpf} and merit further consideration.

The amplituhedron was originally defined as a generalization of both the positive Grassmannian \cite{ArkaniHamed:2012nw} and the polytope description of the NMHV tree amplitude \cite{Hodges:2009hk,ArkaniHamed:2010gg}. More precisely, for the $\text{N}^k$MHV helicity configuration the tree-level amplituhedron is defined as the subspace of $G(k,k{+}4)$, the space of $k$ planes in $(k{+}4)$-dimensions, swept out by positive linear combinations of the positive external data. Triangulating this subspace amounts to finding a non-overlapping set of $(4k)$-dimensional cells in the positive Grassmannian $G_+(k,n)$ covering the full space. This definition of the amplituhedron gives no prescription for actually obtaining a complete collection of such cells, making direct triangulation of the space difficult. This is primarily due to the highly redundant nature of the map from positroid cells to amplituhedra, as the former are always larger than the latter. Recently, an alternative topological definition of the amplituhedron was conjectured \cite{Arkani-Hamed:2017vfh} and verified in many nontrivial cases to be equivalent to the original geometry. In this definition, the amplituhedron is described by a combination of boundary inequalities and a collection of topological sign flip patterns. This new definition gives a completely new and clear understanding of the geometry of the loop-level amplituhedron. For example, in the MHV case, the $\ell$-loop space is decomposed into $\ell$ copies of the one-loop space, together with additional mutual positivity conditions between the different loops. This yields an extremely simple description of the loop-level MHV amplituhedron and (in some cases) makes direct triangulation of the space significantly easier. For example, this new picture, together with an isomorphism between the one-loop MHV and the $m=2,k=2$ amplituhedron, has been utilized to triangulate the two-loop MHV geometry and obtain new representations of the corresponding canonical form \cite{Kojima:2018qzz}. 

Even before tackling the all-loop integrand, there is much to still be understood about the tree-level amplitudes in planar $\mathcal{N}=4$ and the corresponding positive geometries. The beautiful geometric description of the NMHV tree-level space by Hodges \cite{Hodges:2009hk} associates to the amplitude \emph{either} a differential form with logarithmic singularities on the boundaries of a polytope, \emph{or} the volume of the dual polytope, and was foundational to the construction of the amplituhedron. However, the notion of the dual amplituhedron, where ``amplitudes$=$volume'' is literally true, has yet to be made precise for any $k>1$. In this paper, we begin a systematic exploration of the first nontrivial $\text{N}^2$MHV case where the relevant geometry is (in the bosonized $Y$-space) the space of lines in $\mathbb{P}^5$. From the sign flip characterization of the amplituhedron we triangulate the space, obtain a new representation of the canonical form and present the structure of the canonical forms associated to individual sign flip patterns.

The topological definition of the loop-level amplituhedron also leads to novel interpretations of the geometry. For example, the N$^k$MHV one loop amplituhedron can be thought of as the intersection of the $m=2,(k+2)$ and $m=4, k$ tree amplituhedra \cite{Arkani-Hamed:2017tmz, Arkani-Hamed:2017vfh}. Concretely, for $k=1$ the intersection of these two amplituhedra, namely the $m=2, k=3$ amplituhedron and the NMHV tree amplituhedron, is a polygon on a plane. This suggests that the full space can be constructed as a direct product of the subspace spanned by a plane in five dimensions and a point on a polygon. At the level of canonical forms, this corresponds to the product of a degree-six form in the plane and a degree-two form in the point on this polygon. This $6\times2$ description of the one-loop NMHV amplituhedron as a product of two ``unphysical'' $m=2$ amplituhedra is not obvious from the original definition of the amplituhedron or any known amplitudes perspective, where this space is usually described by a $4\times4$ product of the degree-four form in the point $Y$ and degree-four form in the loop line $(AB)$. Importantly, we find the $6\times2$ representation of the space offers a practical advantage to triangulation. 

In this paper, we will obtain novel representations of amplitudes and loop integrands, namely the $\text{N}^2$MHV tree and NMHV tree and one-loop cases, by direct triangulation of the associated amplituhedra from their topological definition. For the NMHV one-loop integrand, our result is directly a product of two $m=2$ amplituhedra, a fact which is obvious from the geometry but is greatly obscured from any other representation of the integrand. Our results, besides being a proof-of-concept of the practical implementation of the amplituhedron technology, offer an intriguing glimpse into the extremely intricate geometry features arising beyond polytopes. 

Our paper is organized as follows. In section \ref{sec:2} we briefly review both definitions of the tree and loop-level amplituhedron and illustrate how to construct the canonical form. In section \ref{sec:3} we solve the NMHV and $\text{N}^2$MHV tree-level triangulation problems by a direct approach. In section \ref{sec:4} we decompose the one-loop NMHV amplituhedron into a product of two $m=2$ amplituhedra, and give an explicit expression of the canonical form of this sign flip representation of the space. In section \ref{sec:5} we discuss the physical interpretation of our results and indicate future directions to extend this work.

\subsection*{Notation}
Scattering amplitudes for $n$ massless particles in planar $\mathcal{N}=4$ super-Yang-Mills are super-functions of the super-momentum twistor variables $(z_a,\tilde{\eta}_a)$, $a=1,\ldots,n$ where the $z_a$ are the momentum twistors of Hodges \cite{Hodges:2009hk}, and the $\tilde{\eta}_a$ are Grassmann variables labelling the helicity configuration. The $Y$-space description of the $m=4$ amplituhedron $\mathcal{A}^{(n,k,\ell)}$ used in this paper involves bosonized twistor variables $Z_a$ which supplement ordinary momentum twistors with $k$ auxiliary Grasssmann parameters $\phi_i^\alpha$, $\alpha=1,\ldots,k$: 
\begin{equation}
Z_a=\begin{pmatrix} z_a \\ \phi_1^A\tilde{\eta}_{1A} \\ \vdots \\ \phi_k^A\tilde{\eta}_{kA} \end{pmatrix},\quad \text{where $A=1,\ldots,4$, for $a=1,\ldots,n$}.
\end{equation}
The bosonized momentum twistors are vectors in a $(k{+}4)$-dimensional projective space. At $\ell$-loop level, the planar integrand can be thought of as a degree-$4\ell$ differential form in the loop variables $\mathcal{L}_1,\ldots,\mathcal{L}_\ell$ which are lines in projective space $\mathbb{P}^3$ (or, equivalently, two-planes in four-dimensions). The canonical differential form $\Omega^{(n,k,\ell)}$ associated to the space $\mathcal{A}^{(n,k,\ell)}$ is a form in a $k$-dimensional plane $Y^I$, $I=1,\ldots,k{+}4$ as well as two-planes in this space $\mathcal{L}_{(i)}^{IJ}=(AB)^{IJ}_{(i)}$, $i=1,\ldots,\ell$, each of which can be represented by the span of two points $A_{(i)}$ and $B_{(i)}$. Alternatively, we can think of $(k{+}2)$-dimensional planes $(YAB)_{(i)}$ all of which intersect on a $k$-plane $Y$. Throughout this work we use the shorthand notation
\begin{equation}
\langle AB\cdots C\rangle=\epsilon_{i_1,\ldots,i_{k{+}4}}A^{i_1}B^{i_2}\cdots C^{i_{k{+}4}}
\end{equation}
to denote contraction with the $(k{+}4)$-dimensional $\epsilon$ tensor. In addition to the $m=4$ space directly relevant for scattering amplitudes, in this paper we frequently consider the $m=2$ amplituhedron $\mathcal{A}^{m=2}_{n,k}$ which lives in $(k{+}2)$-dimensional space \cite{Arkani-Hamed:2017vfh}. 
\section{The amplituhedron}
\label{sec:2}
\subsection{Definition(s) of the amplituhedron}
The original definition of the amplituhedron is a generalization of the interior of plane polygons to the positive Grassmannian \cite{Arkani-Hamed:2013jha}. The tree amplituhedron $\mathcal{A}_{\text{tree}}^{n,k}$ is the space of all $k$-planes $Y^I_\alpha$ in $(k{+}4)$ dimensions which can be written as
\begin{equation}
\label{eq:tree}
Y^I_\alpha =C_{\alpha a}Z^I_a,\qquad \text{for}\quad I{=}1,\ldots,k{+}4, \quad a{=}1,\ldots,n,\quad \alpha{=}1,\ldots,k,
\end{equation}
where $C$ is an element of the positive Grassmannian $G_+(k,n)$ and $Z$ is the collection of external data satisfying the positivity conditions
\begin{equation}
\langle Z_{a_1}\cdots Z_{a_{k{+}4}} \rangle>0\qquad \text{for}\quad  a_1,\cdots<a_{k{+}4}.
\end{equation}
The loop-level amplituhedron $\mathcal{A}^{(n,k,\ell)}$ introduces two-planes $\mathcal{L}_{(i)}$, $i=1,\ldots,\ell$ in the four-dimensional complement of the $k$-plane $Y$. These lines are constrained to be linear combinations of the external data 
\begin{equation}
\mathcal{L}^I_{(i)\alpha}=D_{a\alpha(i)}Z^I_a
\end{equation}
such that $D$ is an element of the positive Grassmannian $G_+(2,n)$. The full amplituhedron $\mathcal{A}^{(n,k,\ell)}$ is the space swept out by all $Y$ and $\mathcal{L}_{(i)}$ of the form
\begin{equation}
Y^I_\alpha =C_{\alpha a}Z^I_a,\ \ \ \ \ \mathcal{L}^I_{(i)\alpha}=D_{a\alpha (i)}Z^I_a
\end{equation}
where we have the additional mutual positivity condition between loops, which demands that all ordered minors of the matrix
\begin{equation}
\left(
    \begin{array}{ccccc}
     D_{(i_1)}\\
     \vdots\\
     D_{(i_\ell)}\\
     C\\
    \end{array}
  \right)
\end{equation}
are positive. 

The recent work \cite{Arkani-Hamed:2017vfh} reformulated the amplituhedron using a purely topological and combinatorial description. This characterization of the geometry is a generalization of the face-centered description of the polytope and uses a collection of inequalities associated to the facets of the polytope to define the space. However, for the amplituhedron simply knowing the codimension-one boundaries of the space is not enough; in this case, additional information about sign flip patterns is needed. The sign flip definition of the tree-level amplituhedron is:\footnote{For concreteness in this paper we choose the sequence $\{\langle Y123i\rangle\}_{i=4,\ldots,n}$ to describe the space, but any other sequence is equivalent.}
\begin{equation}
\begin{split}
&\text{$Y$ is in the $m=4$ amplituhedron iff}\\
&\text{$\langle Yii{+}1jj{+}1\rangle>0$ and the sequence $\{\langle Yabb{+}1i \rangle\}_{i\neq a,b,b{+}1}$ has $k$ sign flips}. 
\end{split}
\end{equation}
The sign flip definition of the loop-level amplituhedron supplements the tree-level conditions with two kinds of conditions: each loop must be in a copy of the one-loop amplituhedron, \emph{and} the loops must be mutually positive. This gives the definition for $\mathcal{A}^{(n,k,\ell)}$ to be the space of all $(k{+}2)$-planes $(YAB)_{\gamma}$, $\gamma=1,\ldots,\ell$ and common $k$-plane $Y$ such that
\begin{equation}
\label{eq:defloop}
\begin{split}
&\langle (YAB)_{\gamma} ii+1 \rangle >0,\ \langle Yii{+}1jj{+}1\rangle>0,\\
&\{\langle (YAB)_\gamma1i \rangle\}_{i=2,\ldots,n}\quad \text{has $k{+}2$ sign flips},\\
&\{\langle Y123i \rangle\}_{i=4,\ldots,n}\text{ has $k$ sign flips},\\
&\langle Y(AB)_\gamma(AB)_\rho\rangle>0.
\end{split}
\end{equation}
The tree-level scattering amplitudes and loop-level integrands of planar $\mathcal{N}{=}4$ sYM are extracted directly from the canonical forms $\Omega^{(n,k,\ell)}$ defined to have logarithmic singularities on all boundaries of $\mathcal{A}^{(n,k,\ell)}$. This form can be written as 
\begin{equation} 
\Omega^{(n,k,\ell)}=\left(\prod_{\alpha=1}^k \langle Y\mathrm{d}^4 Y_\alpha\rangle\right)\left(\prod_{i=1}^\ell \langle (YAB)_i\mathrm{d}^2A_i\rangle\langle (YAB)_i\mathrm{d}^2 B_i\rangle\right)\omega^{(n,k,\ell)},
\end{equation}
where $\omega^{(n,k,\ell)}$ is a rational function in $Y$ and $(YAB)_i$. Throughout this work we will often suppress the measure factors, which should be obvious from context, and therefore do not distinguish between $\Omega$ and $\omega$. To obtain the loop integrand as a form in $\mathbb{P}^3$, one localizes $Y$ and $\mathcal{L}_i=(AB)_i$ to 
\begin{equation}
Y\rightarrow\begin{pmatrix} 0_{4\times k} \\ 1_{k\times k}\end{pmatrix}, \quad \mathcal{L}_i\rightarrow\begin{pmatrix} \tilde{\mathcal{L}}_{i,2\times4}\vert 0_{2\times k},\end{pmatrix}
\end{equation}
and integrates over the four-dimensional Grassmann variables $\phi_1,\ldots,\phi_k$ \cite{Arkani-Hamed:2013jha}.
\section{Sign flip triangulations of tree level amplituhedra}\label{sec:3}
Even at tree level, the geometry of the amplituhedron is well understood only for $k{<}2$.\footnote{Here we refer to the $m=4$ amplituhedron relevant for the scattering amplitudes and loop integrands of planar $\mathcal{N}=4$. For $m=2$ the space has been triangulated for arbitrary $n,k$ \cite{Arkani-Hamed:2017vfh}.} To be clear, for $k=2$ the BCFW recursion does correspond to a geometric triangulation i.e., the regions in $Y$-space corresponding to individual terms are non-overlapping and cover the full space. However, in this case the actual mechanism by which the BCFW recursion (and all other known representations of the amplitude) triangulates the space is unclear. Said differently, starting from the geometry problem it is highly non-trivial to triangulate the space by directly solving the inequalities which define it. Moreover, when this is done, the resulting expression for the canonical form does not, in general, correspond to any particular BCFW (or obvious alternative) type of recursion. This is a radical departure from the $k=1$ geometry, where the original ``$Y=C\cdot Z$'' definition of the amplituhedron naturally leads to the BCFW representation of the NMHV amplitude. However, even in this simple case the sign flip definition is \emph{not} directly associated to individual cells in the positive Grassmannian. Unsurprisingly, triangulations obtained from this characterization of the space in general cannot be identified with individual BCFW terms. To illustrate this distinction, we begin by considering the simple case of the NMHV tree amplitude, where the geometry is understood both from the ``$Y=C\cdot Z$''  picture \cite{Arkani-Hamed:2013jha} as well as the more global description given in~\cite{Arkani-Hamed:2014dca}. From the sign flip definition of the space, we land on a different triangulation distinct from the usual BCFW or CSW recursion. Using the results of this warm-up exercise, we proceed to the $\text{N}^2$MHV case and provide a new triangulation of this amplituhedron.

\subsection{NMHV tree}
The $k=1$ tree-level amplituhedron corresponding to the NMHV tree-level amplitude is labelled by four-dimensional cells of the positive Grassmannian $G_+(1,n)$, and the canonical BCFW recursion triangulates the region as \cite{Arkani-Hamed:2013jha}
\begin{equation}
\label{eq:nmhvBCFWAlln}
\Omega_n^{\text{NMHV}}=\sum_{i<j}[1ii{+}1jj{+}1],
\end{equation}
where the $R$-invariant is defined in $Y$-space as\footnote{In all subsequent expressions in this section we suppress the measure factor.}
\begin{equation}
[abcde]:=\frac{\langle abcde\rangle^4\langle Y\mathrm{d}^4 Y\rangle}{\langle Yabcd\rangle\cdots \langle Yeabc\rangle}.
\end{equation}

\subsubsection{Six point}
 At six points there are three terms in (\ref{eq:nmhvBCFWAlln}),
\begin{equation}\label{eq:6ptbcfw}
\Omega_6^{\text{NMHV}}=[12345]+[12356]+[13456].
\end{equation}
In terms of sign flips, this amplituhedron is comprised of two regions depending on the sign of $\langle Y1235\rangle\lessgtr0$, together with the co-dimension one boundary inequalities $\langle Yii{+}1jj{+}1\rangle>0$ and (accounting for the cyclic symmetry of odd $k$) $\langle Yii{+}1n1\rangle>0$. Parametrizing a generic point $Y$ in terms of five twistors, it is simple to solve the (linear) inequalities defining the regions to obtain the alternate representation of the form\footnote{For the NMHV tree case we omit the derivations of our results. In section~\ref{sec:n2mhv} we illustrate in a more detailed manner the general technique for reducing inequalities and obtaining the corresponding canonical forms.} 
\begin{equation}\label{eq:6ptsignflip}\Omega_6^{\text{NMHV}}=\Omega^{+}+\Omega^{-}, \end{equation} where the individual sign flip patterns have the associated canonical forms
\begin{equation}
\begin{split}
\Omega^{+}&=\frac{\langle12356\rangle\langle13456\rangle^3}{\langle Y1235\rangle\langle Y1346\rangle\langle Y1356\rangle\langle Y1456\rangle\langle Y3456\rangle}+[12356],\\
\Omega^{-}&=-\frac{\langle 12345\rangle\langle13456\rangle^3}{\langle Y1235\rangle\langle Y1345\rangle\langle Y1346\rangle\langle Y1456\rangle\langle Y3456\rangle}+[12345].
\end{split}
\end{equation}
To verify that (\ref{eq:6ptbcfw}) and (\ref{eq:6ptsignflip}) agree is a straightforward algebraic exercise using the Schouten identity. We see the decomposition (\ref{eq:6ptsignflip}) can alternatively be thought of as breaking the amplituhedron into four regions:
\begin{align}
\begin{split}
r_1 & =\{\langle Y1235\rangle>0,\langle Y3461\rangle>0,\langle Y1356\rangle>0,\langle Y4561\rangle>0,\langle Y3456\rangle>0\},\\
r_2 & =\{\langle Y1235\rangle>0,\langle Y2356\rangle>0,\langle Y1356\rangle<0,\langle Y1256\rangle>0,\langle Y2361\rangle>0\},\\
r_3 & =\{\langle Y1235\rangle<0,\langle Y1345\rangle>0,\langle Y3461\rangle>0,\langle Y4561\rangle>0,\langle Y3456\rangle>0\},\\
r_4 & =\{\langle Y1234\rangle>0,\langle Y2345\rangle>0,\langle Y1345\rangle<0,\langle Y1245\rangle>0,\langle Y1235\rangle<0\}.
\end{split}
\end{align}
Here the regions $r_2$ and $r_4$ correspond to the $R$-invariants $[12356]$ and $[12345]$, respectively, while the union of $r_1$ and $r_3$ is exactly the region whose canonical form is $[13456]$. We see that in order to compute the forms it is natural to impose definite signs on the brackets $\langle Y1345\rangle$ and $\langle Y1356\rangle$ (which cancel in the sum) to make the individual building blocks simple. For higher $n$ and $k$ the complexity of the canonical form for each sign flip pattern increases, and computing them requires cutting the regions with additional inequalities -- in fact, choosing these conditions expediently can dramatically simplify the individual pieces in the triangulation.

\subsubsection{Seven point}
It is instructive to continue our NMHV example with the seven point case. Here, the region corresponding to the amplitude is given by three sign patterns, so the form can be written as
\begin{equation}
\Omega_7^{\text{NMHV}}=\Omega^{++}+\Omega^{+-}+\Omega^{--}.
\end{equation}
The canonical form for the all-minus sign pattern is extremely simple:
\begin{align}
\Omega^{--}=&[12345]+\frac{\langle12345\rangle\langle34567\rangle^3}{\langle Y1235\rangle\langle Y3456\rangle\langle Y3457\rangle\langle Y3467\rangle\langle Y4567\rangle}\nonumber\\&+\frac{\langle12345\rangle\langle34571\rangle^3}{\langle Y1235\rangle\langle Y3457\rangle\langle Y3451\rangle\langle Y3471\rangle\langle Y4571\rangle}.
\end{align}
Geometrically this splits the region into three pieces 
\begin{align}
r^{--}=&\{\langle Y1234\rangle>0,\langle Y2345\rangle>0,\langle Y1345\rangle<0,\langle Y1245\rangle>0,\langle Y1235\rangle<0\}\nonumber\\
&\cup\{\langle Y1235\rangle<0,\langle Y3456\rangle>0,\langle Y3457\rangle<0,\langle Y3467\rangle>0,\langle Y4567\rangle>0\}\nonumber\\
&\cup\{\langle Y1235\rangle<0,\langle Y1345\rangle>0,\langle Y3457\rangle>0,\langle Y3471\rangle>0,\langle Y4571\rangle>0\}.
\end{align}
We see that once the signs $\langle Y1235\rangle$ and $\langle Y1236\rangle$ are fixed to be negative, further specifying the signs of the sequence $\{\langle Y345i\rangle\}_{i\neq3,4,5}$ gives a natural triangulation of the region. The expression~(\ref{eq:allMinus7}) can be written more suggestively as
\begin{equation}
\label{eq:allMinus7}
\Omega^{--}=\sum_{i=6,7,1}\frac{\langle12345\rangle\langle345ii{+}1\rangle^3}{\langle Y1235\rangle\langle Y345i\rangle\langle Y345i{+}1\rangle\langle Y34ii{+}1\rangle\langle Y45ii{+}1\rangle}.
\end{equation}
This expression also implies the co-dimension one surface $\langle Y1235\rangle{=}0$ is a boundary of this sign pattern, while $\langle Y1236\rangle{=}0$ is not. 

Similarly, for the all-plus sign pattern the region can be triangulated by imposing definite signs on the additional sequence $\{\langle Y671i\rangle\}_{i\neq6,7,1}$, with the decomposition of the form being 
\begin{align}
\Omega^{{+}{+}}=\sum_{i=2,3,4}\frac{(-1)\langle12367\rangle\langle671ii{+}1\rangle^3}{\langle Y1236\rangle\langle Y671i\rangle\langle Y671i{+}1\rangle\langle Y67ii{+}1\rangle\langle Y71ii{+}1\rangle}.
\end{align}
For the $(+-)$ sign pattern, both $\langle Y1235\rangle{=}0$ and $\langle Y1236\rangle{=}0$ co-dimension one boundaries are accessible, and the decomposition of the form is 
\begin{align}
\label{eq:pm}
\Omega^{+-}=&\sum_{i=1,2,3}\frac{(-1)\langle12356\rangle\langle567ii{+}1\rangle^3}{\langle Y1235\rangle\langle Y567i\rangle\langle Y567i{+}1\rangle\langle Y56ii{+}1\rangle\langle Y67ii{+}1\rangle}\nonumber\\&+\sum_{i=2,3,4}\frac{\langle Y123(56){\cap}(7123)\rangle\langle671ii{+}1\rangle^3}{\langle Y1235\rangle\langle Y1236\rangle\langle Y671i\rangle\langle Y671i{+}1\rangle\langle Y67ii{+}1\rangle\langle Y71ii{+}1\rangle}.
\end{align}
By expanding the intersection in the numerator of the second term in (\ref{eq:pm}) in the sum of the three forms the terms containing the pole $\langle Y1236\rangle$ cancel, and the canonical form for the full amplitude is 
\begin{equation}
\Omega_7^{\text{NMHV}}{=}\sum_{\substack{j\neq1,2,3,5 \\ i\neq j{-}2,j{-}1,j,j{+}1}}\frac{\langle1235j\rangle\langle j{-}1jj{+}1ii{+}1\rangle^3}{\langle Y1235\rangle\langle Yj{-}1jj{+}1i\rangle\langle Yj{-}1jj{+}1i{+}1\rangle\langle Yj{-}1jii{+}1\rangle\langle Yjj{+}1ii{+}1\rangle}.
\end{equation}
\subsubsection{All multiplicity generalization}
At $n$ points, the NMHV tree amplituhedron decomposes into $n{-}5$ regions which can be labelled by the number of brackets in the list $\langle Y1235\rangle,\ldots,\langle Y123n{-}1\rangle$ which are negative. For these sign patterns, the key bracket is that which labels the flip from positive to negative. For example, for the all-minus sign pattern where $\langle Y123i\rangle<0$ for $i=5,\ldots,n$, the form is given by the obvious generalization of the seven point case (\ref{eq:allMinus7}),
\begin{equation}
\Omega^{-,\ldots,-}=\sum_{i\neq2,3,4,5}\frac{\langle12345\rangle\langle345ii{+}1\rangle^3}{\langle Y1235\rangle\langle Y345i\rangle\langle Y345i{+}1\rangle\langle Y34ii{+}1\rangle\langle Y45ii{+}1\rangle},
\end{equation}
where $\langle Y1235\rangle$ is the first bracket in the sequence which flips sign and hence is present in all terms, and we have used the brackets $\langle Y345i\rangle$ to further cut the region into elementary pieces. 

Just as at seven points, in the sum over all sign patterns all terms with poles $\langle Y1236\rangle,\ldots,\langle Y123n{-}1\rangle$ cancel, leaving the result
\begin{equation}
\label{eq:nmhvAlln}
\Omega_n^{\text{NMHV}}{=}\sum_{i,j}\frac{\langle1235j\rangle\langle j{-}1jj{+}1ii{+}1\rangle^3}{\langle Y1235\rangle\langle Yj{-}1jj{+}1i\rangle\langle Yj{-}1jj{+}1i{+}1\rangle\langle Yj{-}1jii{+}1\rangle\langle Yjj{+}1ii{+}1\rangle},
\end{equation}
which matches the BCFW representation of the $n$-point amplitude (\ref{eq:nmhvBCFWAlln}). Note that the choice $(1235)$ is completely arbitrary and can be replaced by any three-plane. 

\subsection{$\text{N}^2$MHV tree}\label{sec:n2mhv}
In the rest of this section, we take as our starting point the sign flip characterization of the $\text{N}^2$MHV tree amplituhedron, which we repeat here: we consider the space of all two-planes $Y$ in six dimensions satisfying
\begin{equation}
\begin{split}
&\langle Yii{+}1jj{+}1\rangle>0,\quad \langle Y1ii{+}1n\rangle>0,\\
&\{\langle Y123i\rangle\}_{i=4,\ldots,n}\quad \text{has 2 sign flips}.
\end{split}
\end{equation}
We begin with the cases where $n=6,7$ and then generalize to all multiplicity.
\subsubsection{Six point}\label{sec:n2mhv6}
The first case where the $k=2$ helicity configuration yields a nonzero tree-level amplitude is $n=6$. At this multiplicity, the $\text{N}^2$MHV tree-level amplituhedron corresponds to a single sign pattern
\begin{equation}
\label{eq:sixPtIneq}
A_6^{\text{N}^2\text{MHV}}=\{\langle Yii{+}1jj{+}1\rangle>0,\langle Y1ii{+}16\rangle>0,\langle Y1235\rangle<0\}.
\end{equation}
Written as an eight-form in $Y$-space, the amplitude is given by a single term
\begin{equation}
\label{eq:n2mhv6}
\Omega_6^{\text{N}^2\text{MHV}}=\frac{\langle123456\rangle^4\langle Y\mathrm{d}^4Y_1\rangle\langle Y\mathrm{d}^4 Y_2\rangle}{\langle Y1234\rangle\langle Y2345\rangle\langle Y3456\rangle\langle Y4561\rangle\langle Y5612\rangle\langle Y6123\rangle}.
\end{equation}
In this case, the sign flip characterization of the space gives no advantage in reproducing this result over any previous method -- the geometry is too simple. However, to illustrate the basic tools and notation needed for the higher multiplicity case considered below we will be explicit in this warm-up exercise of re-deriving (\ref{eq:n2mhv6}). 

A two-plane in six dimensions is equivalent to a line in the projective space $\mathbb{P}^5$ and generically has eight degrees of freedom. If we parametrize this line as the span of two points, $Y=(Y_1Y_2)$, then $GL(2)$ invariance allows us to write without loss of generality
\begin{equation}
\begin{split}
Y_1&=Z_1+\alpha_1Z_2+\alpha_2Z_3+\alpha_3Z_4+\alpha_4Z_5,\\
Y_2&=Z_2+\beta_1Z_3+\beta_2Z_4+\beta_3Z_5+\beta_4Z_6.
\end{split}
\end{equation}
In this parametrization the projective measure is $\langle Y\mathrm{d}^4Y_1\rangle\langle Y\mathrm{d}^4 Y_2\rangle=\langle123456\rangle^2\beta_4\mathrm{d}^4\alpha\mathrm{d}^4\beta$, and the map back to $Y$-space can be written as
\begin{equation}
\label{eq:map6}
\begin{split}
\alpha_1&=-\frac{\langle Y1345\rangle}{\langle Y2345\rangle},\quad \alpha_2=\frac{\langle Y1245\rangle}{\langle Y2345\rangle},\quad \alpha_3=-\frac{\langle Y1235\rangle}{\langle Y2345\rangle},\quad \alpha_4=\frac{\langle Y1234\rangle}{\langle Y2345\rangle},\\
\beta_1&=-\frac{\langle Y2456\rangle}{\langle123456\rangle},\quad \beta_2=\frac{\langle Y2356\rangle}{\langle 123456\rangle},\quad \beta_3=-\frac{\langle Y2346\rangle}{\langle 123456\rangle},\quad \beta_4=\frac{\langle Y2345\rangle}{\langle 123456\rangle}.
\end{split}
\end{equation}
At six points we can set the (six-dimensional) extended external data to the $6{\times}6$ identity matrix by a $GL(6)$ transformation. In this parametrization the only kinematical invariant is set to unity i.e., $\langle 123456\rangle=1$. The inequalities (\ref{eq:sixPtIneq}) are equivalent to
\begin{equation}
A_6^{\text{N}^2\text{MHV}}=\left\{\alpha_1,\alpha_2,\alpha_3,\alpha_4,\beta_4>0, \beta_1>\frac{\alpha_2}{\alpha_1},\beta_2>\frac{\alpha_3\beta_1}{\alpha_2},\beta_3>\frac{\alpha_4\beta_2}{\alpha_3}\right\},
\end{equation}
and the associated logarithmic form is 
\begin{equation}
\Omega_6^{\text{N}^2\text{MHV}}=\frac{\mathrm{d}^4\alpha\mathrm{d}^4\beta}{\alpha_4(\alpha_1\beta_1-\alpha_2)(\alpha_2\beta_2-\alpha_3\beta_1)(\alpha_3\beta_3-\alpha_4\beta_2)\beta_4},
\end{equation}
which matches the amplitude (\ref{eq:n2mhv6}) when written projectively using (\ref{eq:map6}). Note that the factor $\langle 123456\rangle^4$ is required by demanding that the form be projective in each momentum twistor $Z_i$ (as well as the line $Y$).

\subsubsection{Seven point}
For $n=7$ the amplituhedron is the union of three connected regions defined by the signs of the brackets $\langle Y1235\rangle$ and $\langle Y1236\rangle$:
\begin{equation}\label{eq:geometry7}
\mathrm{sign}(\langle Y1235\rangle,\langle Y1236\rangle)=\{(+,-),(-,+),(-,-)\},
\end{equation}
where we leave the inequalities $\langle Yii{+}1jj{+}1\rangle>0$ implicit. Our objective is to compute the three canonical forms $\Omega^{+-},\Omega^{-+}$ and $\Omega^{--}$. Despite the fact that the seven point $\text{N}^2$MHV amplitude is the parity conjugate of the NMHV amplitude, from the geometry perspective this space is already quite nontrivial. The canonical form we aim to reproduce can be obtained by writing the superamplitude obtained from, for example, BCFW recursion in the $Y$-space of the amplituhedron. The result is a sum of six terms
\begin{align} \label{eq:bcfwresult}  \Omega_7^{\text{N}^2\text{MHV}}=&\frac{\langle 123456\rangle^4}{\begin{array}{c}\langle Y1234\rangle\langle Y2345\rangle\langle Y3456\rangle\\\textcolor{hblue}{\langle Y1456\rangle}\langle Y1256\rangle\textcolor{hblue}{\langle Y1236\rangle}\end{array}}{+}\frac{\langle 134567\rangle^4}{\begin{array}{c}\textcolor{hblue}{\langle Y1345\rangle}\langle Y3456\rangle\langle Y4567\rangle\\\langle Y1567\rangle\textcolor{hblue}{\langle Y1367\rangle}\langle Y1347\rangle\end{array}}\nonumber\\{+}&\frac{\langle123467\rangle^4}{\begin{array}{c}\langle Y1234\rangle\textcolor{hblue}{\langle Y2346\rangle}\langle Y3467\rangle\\\textcolor{hblue}{\langle Y1467\rangle}\langle Y1267\rangle\langle Y1237\rangle\end{array}}{+}\frac{\langle Y(12367){\cap}(14567)\rangle^4}{\begin{array}{c}\langle Y1237\rangle\langle Y1267\rangle\textcolor{hblue}{\langle Y1367\rangle}\langle Y4567\rangle\textcolor{hblue}{\langle Y1467\rangle}\langle Y1567\rangle\\ \textcolor{hblue}{\langle Y(45)\cap(Y623)671\rangle}\textcolor{hblue}{\langle Y(45)\cap(Y123)671\rangle}\end{array}}\nonumber\\ {+}&\frac{\langle Y(23456){\cap}(12367)\rangle^4}{\begin{array}{c}\langle Y2345\rangle\langle Y3456\rangle\textcolor{hblue}{\langle Y2346\rangle}\langle Y2356\rangle\textcolor{hblue}{\langle Y1236\rangle}\langle Y1267\rangle\\ \langle Y2367\rangle \langle Y1237\rangle\textcolor{hblue}{\langle Y(45)\cap(Y623)671\rangle}\end{array}}\nonumber\\&{+}\frac{\langle Y(12345)\cap(14567)\rangle^4}{\begin{array}{c}\langle Y1234\rangle\langle Y1245\rangle\textcolor{hblue}{\langle Y1345\rangle}\langle Y2345\rangle\textcolor{hblue}{\langle Y1456\rangle}\langle Y1567\rangle\\ \langle Y4567\rangle\langle Y1457\rangle\textcolor{hblue}{\langle Y(45)\cap(Y123)671\rangle}\end{array}}.\end{align}
The BCFW representation triangulates the amplituhedron internally by introducing spurious poles which cancel pairwise (we indicate spurious poles by \textcolor{hblue}{blue} text in the above expressions). An alternative representation of the canonical form is given by the CSW expansion involving an arbitrary point $Z_\star$, which is more useful for the all-$n$ comparison in section~\ref{sec:n2mhvAlln}. However, this decomposition of the amplitude does not triangulate the space in the usual (i.e., internal) sense. Demonstrating the equivalence of these forms is an extremely nontrivial algebraic exercise (as is actually canceling the spurious poles in (\ref{eq:bcfwresult})). Furthermore, from the sign flip definition of the amplituhedron neither of these representations seem natural, and it is difficult to identify some combination of terms in (\ref{eq:bcfwresult}) as corresponding to any particular sign pattern. In this section our goal is to reproduce the canonical form starting directly from the sign flip definition of the geometry (\ref{eq:geometry7}). As we shall see, this gives a drastically different representation of the amplitude and suggests a novel collection of all-$n$ geometries directly associated to individual sign flip patterns. 

At seven points, the parametrization for $Y$ used in section~\ref{sec:n2mhv6} is not ideal as it leads to a significant number of quadratic inequalities. Instead, we utilize a parametrization involving all seven momentum twistors:
\begin{equation}
\begin{split}
Y_1&=Z_1+\alpha_1Z_2+\alpha_2Z_3+\alpha_3Z_4+\alpha_4Z_5,\\
Y_2&=Z_3+\beta_1Z_4+\beta_2Z_5+\beta_3Z_6+\beta_4Z_7.
\end{split}
\end{equation}
In this set of coordinates, the projective measure is 
\begin{equation} 
\langle Y\mathrm{d}^4Y_1\rangle\langle Y\mathrm{d}^4 Y_2\rangle=\mathrm{d}^4\alpha \mathrm{d}^4\beta \left(\beta_3\langle123456\rangle+\beta_4\langle123457\rangle\right)\left(\langle134567\rangle+\alpha_1\langle234567\rangle\right),\end{equation} 
and the map back to projective coordinates can be written as
\begin{equation}
\begin{split}
\alpha_1&=-\frac{\langle Y1345\rangle}{\langle Y2345\rangle},\quad \alpha_2=\frac{\langle Y1245\rangle}{\langle Y2345\rangle},\quad \alpha_3=-\frac{\langle Y1235\rangle}{\langle Y2345\rangle},\quad \alpha_4=\frac{\langle Y1234\rangle}{\langle Y2345\rangle},\\
\beta_1&=-\frac{\langle Y3567\rangle}{\langle Y4567\rangle},\quad \beta_2=\frac{\langle Y3467\rangle}{\langle Y4567\rangle},\quad \beta_3=-\frac{\langle Y3457\rangle}{\langle Y4567\rangle},\quad \beta_4=\frac{\langle Y3456\rangle}{\langle Y4567\rangle}.
\end{split}
\end{equation}
At seven points the amplituhedron is cut out by the combination of the fourteen co-dimension one boundaries $\langle Yii{+}1jj{+}1\rangle>0$, 
\begin{equation} 
\begin{split}
\label{eq:ineq1}
\{&\langle Y1234\rangle>0,\langle Y1245\rangle>0,\langle Y1256\rangle>0,\langle Y1267\rangle>0,\\ &\langle Y2345\rangle>0,\langle Y2356\rangle>0,\langle Y2367\rangle>0,\langle Y1237\rangle>0,\\ &\langle Y3456\rangle>0,\langle Y3467\rangle>0,\langle Y1347\rangle>0,\langle Y4567\rangle>0,\\ &\langle Y1457\rangle>0,\langle Y1567\rangle>0\},
\end{split}
\end{equation}
 along with the sign patterns of the $\{\langle Y123i\rangle\}_{i\neq1,2,3}$ sequence (\ref{eq:geometry7}). Imposing these constraints defines a system of linear and quadratic constraints on the eight parameters $\alpha_1,\ldots,\beta_4$; we seek the canonical form with logarithmic singularities on all boundaries of this space. To make the calculation simpler, without loss of generality we fix the external data at seven points to be
\begin{equation}
Z=(Z_1Z_2Z_3Z_4Z_5Z_6Z_7)=\begin{pmatrix} \mqty{\imat{6}} & \mqty{1 \\ -1 \\ 1 \\ -1 \\ 1 \\ -1} \end{pmatrix},
\end{equation}
which sets $\langle abcdef\rangle{=}1$ for $a<b<c<d<e<f$, thereby trivializing the positivity constraints on the external data. One can alternatively replace the $\pm1$ in $Z_7$ by any numbers of alternating sign to also satisfy the constraints; we make this choice for simplicity. To compute the canonical form associated to a set of inequalities essentially amounts to finding the full-dimensional component of a \emph{cylindrical algebraic decomposition} (CAD) \cite{Collins:1976} of this semi-algebraic set, which divides it into disjoint cells described by so-called ``cylindrical'' conditions. In our case, this means that each cell is described in the variables $\alpha_1,\ldots,\beta_4$ by inequalities of the form
\begin{equation}
\{a_1<\alpha_1<b_1,a_2(\alpha_1)<\alpha_2<b_2(\alpha_1),\ldots,a_8(\alpha_1,\ldots,\beta_3)<\beta_4<b_8(\alpha_1,\ldots,\beta_3)\},
\end{equation}
where in this ordering $a_1,b_1$ are constants, $a_2,b_2$ can only depend on $\alpha_1$, etc. From this description it is straightforward to write the associated logarithmic form by repeatedly using the fact that
\begin{equation}
a<\alpha<b\quad \text{has the canonical form} \quad \frac{(b-a)\mathrm{d}\alpha}{(\alpha-a)(b-\alpha)}.
\end{equation}
While obtaining a CAD for arbitrary semi-algebraic sets is in principle always possible, the computational complexity even in the case of quadratic inequalities is well-known to be doubly exponential in the number of variables, making this a nontrivial task.\footnote{In \textsc{Mathematica} efficient CAD algorithms are implemented with the built-in functions \texttt{Reduce} and (the significantly faster) \texttt{GenericCylindricalDecomposition}. However, the amount of time required to solve nontrivial sets of inequalities relevant for scattering amplitudes is highly dependent on both the parametrization used and the ordering of variables, and requires significant patience.} 

A well-defined approach in this problem is to initially impose a subset of the inequalities (\ref{eq:geometry7}) and (\ref{eq:ineq1}), triangulate this intermediate result into elementary pieces, then further cut each sub-region with the remaining inequalities (one at a time, if need be). This method of calculation involves choosing the initial set of inequalities to impose, the order of the remaining inequalities used to further divide the sub-regions and the ordering of the variables. Each of these choices can greatly impact both the total number and complexity of each sub-region. After all inequalities have been imposed, in order to discern $n$-point structure it is generically necessary to post-process the list of sub-regions. This involves cutting some sub-regions further to make the canonical forms simpler, as well as combining different sub-regions to cancel unnecessary spurious poles (thus also simplifying the canonical form). To obtain a seven-point result which leads directly to an all-multiplicity generalization, we seek to simultaneously optimize both the number of terms in the cell decomposition and the complexity of the canonical forms corresponding to individual terms. In this particular example, although the amplitude is given by (\ref{eq:bcfwresult}), the canonical forms corresponding to individual sign flip patterns are a priori unknown. Thus, we begin our analysis by computing these forms in some (possibly non-ideal) representation, and subsequently examine their structure and attempt to write down analogues of the simple NMHV expressions (\ref{eq:nmhvAlln}).

Let us illustrate this approach in detail for the $(-+)$ sign pattern, where $\langle Y1235\rangle<0$ and $\langle Y1236\rangle>0$. If we begin by imposing the inequalities 
\begin{equation}
\begin{split}
\{&\langle Y1234\rangle>0,\langle Y1245\rangle>0,\langle Y1256\rangle>0,\langle Y1267\rangle>0,\langle Y1237\rangle>0,\langle Y2345\rangle>0,\\&\langle Y2356\rangle>0,\langle Y3456\rangle>0,\langle Y3467\rangle>0,\langle Y4567\rangle>0,\langle Y1235\rangle<0,\langle Y1236\rangle>0\},
\end{split}
\end{equation}
using the ordering $\alpha_1,\ldots,\beta_4$ the associated CAD is a list of twenty-four sub-regions. Imposing the remaining four inequalities $\langle Y2367\rangle>0,\langle Y1347\rangle>0,\langle Y1457\rangle>0$ and $\langle Y1567\rangle>0$ reduces this to a list of seven regions. For example, in our parametrization one region is 
\begin{equation}
\label{eq:reg1}
\{\alpha_1,\alpha_2,\alpha_3,\alpha_4>0,\beta_1>\frac{\alpha_3}{\alpha_2},\beta_2>\frac{\alpha_4\beta_1}{\alpha_3},\beta_3>0,0<\beta_4<\frac{\alpha_3\beta_2-\alpha_4\beta_1}{\alpha_3+\alpha_4}\},
\end{equation}
which has the canonical form\footnote{Note that in this expression and all that follow in this section, we suppress measure factors.}
\begin{equation}
\omega_1^{-+}=\frac{1}{\alpha_1\alpha_4(\alpha_3-\alpha_2\beta_1)\beta_3\beta_4(\alpha_4\beta_1-\alpha_3\beta_2+\alpha_3\beta_4+\alpha_4\beta_4)},
\end{equation}
or written projectively
\begin{equation}
\label{eq:ex7}
\omega_1^{-+}{=}\frac{-\langle123456\rangle\langle Y(12345){\cap}(34567)\rangle^3}{\langle Y1234\rangle\langle Y1236\rangle\langle Y1345\rangle\langle Y2345\rangle\langle Y3456\rangle\langle Y3457\rangle\langle Y4567\rangle\langle Y(12(34){\cap}(Y567)5\rangle},
\end{equation}
where $\langle Y(12(34){\cap}(Y567)5\rangle{=}\langle Y1245\rangle\langle Y3567\rangle{-}\langle Y1235\rangle\langle Y4567\rangle$ (note that this quadratic pole is invariant under relabelling $(12)\leftrightarrow(34)$). A projective description of the region equivalent to (\ref{eq:reg1}) is 
\begin{equation}
\begin{split} \{&\langle Y1234\rangle>0,\langle Y1235\rangle<0,\langle Y1236\rangle>0,\langle Y1345\rangle<0,\langle Y2345\rangle>0,\\ &\langle Y3456\rangle>0,\langle Y3457\rangle<0,\langle Y4567\rangle>0,\langle Y(12(34){\cap}(Y567)5\rangle>0\}.\end{split} 
\end{equation}
Note that although $\langle Y1235\rangle$ is not a pole of the form (or equivalently a co-dimension one boundary of the region), its sign is still required to be fixed (and is not implied by the other conditions). This is a generic feature of positive geometries and can be seen even in the local integral representation of the MHV one-loop amplitude which externally triangulates the amplituhedron \cite{Langer:2020}. Just as at seven points, the factor $\langle123456\rangle$ in (\ref{eq:ex7}) is needed to restore projectivity of the form in $Z_i$ -- although all brackets are set to unity in our choice of external data, at seven points it is always trivial to restore such factors from this requirement. For conciseness in all subsequent seven point canonical forms (\ref{eq:forms1})--(\ref{eq:forms2}) we introduce the shorthand notation $[a]=\langle bcdefg\rangle$ where $a$ is the element of the set $\{1,\ldots,7\}$ which is not $b,c,d,e,f$ or $g$. Three pairs of the remaining six CAD sub-regions individually combine to give a total of three additional building blocks to complete the $(-+)$ sign pattern space:
\begin{equation}
\Omega^{-+}=\sum_{i=1}^4 \omega_i^{-+},
\end{equation}
where
\begin{align}
\label{eq:forms1}
\omega_2^{-+}&=\frac{-[7]\langle Y(12345){\cap}(12567)\rangle^3}{\begin{array}{c}\langle Y1234\rangle\langle Y1236\rangle\langle Y1245\rangle\langle Y1256\rangle\langle Y1257\rangle\langle Y1567\rangle\langle Y2345\rangle\langle Y(12(34){\cap}(Y567)5\rangle\end{array}},\\
\omega_3^{-+}&=\frac{[7][6]^3}{\langle Y1234\rangle\langle Y1236\rangle\langle Y1257\rangle\langle Y1457\rangle\langle Y2345\rangle\langle Y3457\rangle},\\
\omega_4^{-+}&=\frac{[7][2]^3}{\langle Y1236\rangle\langle Y1345\rangle\langle Y1347\rangle\langle Y1567\rangle\langle Y3456\rangle\langle Y4567\rangle}.
\end{align}
Examining these five terms we can see the dissimilarity with the usual BCFW representation. Of course, this decomposition is far from unique or canonical -- the order in which we imposed the inequalities defining the region dramatically affect the representation of the form we obtain. However, given \emph{any} particular representation of the form, it is straightforward to verify that the form can simply be written as 
\begin{equation}
\Omega^{-+}{=}\frac{\mathcal{N}^{-+}}{\begin{array}{c}\langle Y1234\rangle\langle Y1236\rangle\langle Y1245\rangle\langle Y1256\rangle\langle Y1347\rangle\\\langle Y1457\rangle\langle Y1567\rangle\langle Y2345\rangle\langle Y3456\rangle\langle Y4567\rangle\end{array}}.
\end{equation}

Repeating this procedure for the $(+-)$ sign pattern we find eight basic building blocks describe the region (in one particular way of solving the inequalities):
\begin{align}
\label{eq:pm7}
\omega_1^{+-}&{=}\frac{[4][1]^2\left(\begin{array}{c}\langle Y(12345){\cap}(34567)\rangle\langle Y(12345){\cap}(12367)\rangle\\\times\langle Y(12345){\cap}(12(45){\cap}(Y367)67)\rangle\end{array}\right)}{\begin{array}{c}\langle Y1235\rangle\langle Y1237\rangle\langle Y1267\rangle\langle Y1345\rangle\langle Y2367\rangle\langle Y3456\rangle\langle Y3567\rangle\langle Y4567\rangle\\\langle Y(12345){\cap}(24567)\rangle\langle Y(12345){\cap}(23467)\rangle\end{array}},\\
\omega_2^{+-}&{=}\frac{[4]^4}{\langle Y1235\rangle\langle Y1237\rangle\langle Y1267\rangle\langle Y1567\rangle\langle Y2356\rangle\langle Y3567\rangle},\\
\omega_3^{+-}&{=}\frac{[5]^2[4]\langle Y(12345){\cap}(34567)\rangle^2\langle Y(12347){\cap}(34567)\rangle}{\begin{array}{c}\langle Y1235\rangle\langle Y1237\rangle\langle Y1347\rangle\langle Y3456\rangle\langle Y3467\rangle\langle Y4567\rangle\\\langle Y(12345){\cap}(23467)\rangle\langle Y(12345){\cap}(12(34){\cap}(Y567)67)\rangle\end{array}},\\
\omega_4^{+-}&{=}\frac{[4][3]^2\langle Y(12345){\cap}(34567)\rangle\langle Y(12345){\cap}(12367)\rangle\langle Y(12367){\cap}(34567)\rangle}{\begin{array}{c}\langle Y1235\rangle\langle Y1237\rangle\langle Y1267\rangle\langle Y1367\rangle\langle Y3456\rangle\langle Y3567\rangle\langle Y4567\rangle\\ \langle Y(12345){\cap}(24567)\rangle\langle Y(12345){\cap}(12467)\rangle\end{array}},\\
\omega_5^{+-}&{=}\frac{[5][4][2]\langle Y(12345){\cap}(13467)\rangle\langle Y(12345){\cap}(34567)\rangle\langle Y(12567){\cap}(34567)\rangle}{\begin{array}{c}\langle Y1235\rangle\langle Y1347\rangle\langle Y1367\rangle\langle Y1567\rangle\langle Y3456\rangle\langle Y4567\rangle\\ \langle Y(12345){\cap}(23467)\rangle\langle Y(12345){\cap}(12(34){\cap}(Y567)67)\rangle\end{array}},\\
\omega_6^{+-}&{=}\frac{[4][2]^2[1]\langle Y(12345){\cap}(13467)\rangle}{\langle Y1235\rangle\langle Y1347\rangle\langle Y1367\rangle\langle Y1567\rangle\langle Y3456\rangle\langle Y4567\rangle\langle Y(12345){\cap}(23467)\rangle},\\
\omega_7^{+-}&{=}\frac{-[4][2][1]^2\left(\begin{array}{c}\langle Y(12345){\cap}(12367)\rangle\langle Y123(45){\cap}(Y367)\rangle\\\times\langle Y(12345){\cap}(12(45){\cap}(Y367)67)\rangle\end{array}\right)}{\begin{array}{c}\langle Y1235\rangle\langle Y1237\rangle\langle Y1267\rangle\langle Y1345\rangle\langle Y1367\rangle\langle Y2367\rangle\langle Y3456\rangle\\\langle Y3567\rangle\langle Y4567\rangle\langle Y(12345){\cap}(24567)\rangle\langle Y(12345){\cap}(23467)\rangle\end{array}},\\
\omega_8^{+-}&{=}\frac{-[5]^2[4]\left(\begin{array}{c}\langle Y(12345){\cap}(34567)\rangle\langle Y(12345){\cap}(12567)\rangle\\\times\langle Y(12345){\cap}(12367)\rangle\langle Y(12367){\cap}(34567)\rangle\end{array}\right)}{\begin{array}{c}\langle Y1235\rangle\langle Y1237\rangle\langle Y1267\rangle\langle Y1367\rangle\langle Y3456\rangle\langle Y3567\rangle\langle Y(12345){\cap}(12467)\rangle\\\langle Y(12345){\cap}(23467)\rangle \langle Y(12345){\cap}(12(34){\cap}(Y567)67)\rangle\end{array}}.
\end{align}
For this sign pattern, by a simple residue check one finds the following structure for the canonical form: 
\begin{equation}
\Omega^{+-}{=}\frac{\mathcal{N}^{+-}}{\begin{array}{c}\langle Y1235\rangle\langle Y1237\rangle\langle Y1267\rangle\langle Y1347\rangle\langle Y1567\rangle\\\langle Y2356\rangle\langle Y2367\rangle\langle Y3456\rangle\langle Y3467\rangle\langle Y4567\rangle\end{array}}.
\end{equation}
Finally, for the $(--)$ sign pattern one possible decomposition of the form is:
\begin{align}
\omega_1^{--}&{=}\frac{\langle Y(12345){\cap}(12367)\rangle\langle Y(12345){\cap}(14567)\rangle^3\langle Y(12345){\cap}(34567)\rangle}{\begin{array}{c}([2]\langle Y1245\rangle\langle Y123(45){\cap}(Y367)\rangle{+}\langle Y1345\rangle\langle Y(12(45){\cap}(Y123)67){\cap}(34567)\rangle)\\ \langle Y1234\rangle\langle Y1236\rangle\langle Y1245\rangle\langle Y1345\rangle\langle Y1457\rangle\langle Y1567\rangle\langle Y2345\rangle\langle Y4567\rangle\end{array}},\\
\omega_2^{--}&{=}\frac{-\langle Y(12345){\cap}(12367)\rangle^4\langle Y(12345){\cap}(34567)\rangle}{\begin{array}{c}([2]\langle Y1245\rangle\langle Y123(45){\cap}(Y367)\rangle{+}\langle Y1345\rangle\langle Y(12(45){\cap}(Y123)67){\cap}(34567)\rangle) \\ \langle Y1234\rangle\langle Y1235\rangle\langle Y1236\rangle\langle Y1237\rangle\langle Y1267\rangle\langle Y2345\rangle \langle Y123(45){\cap}(Y367)\rangle \end{array}},\\
\omega_3^{--}&{=}\frac{-[2]^3\langle Y(12345){\cap}(12367)\rangle\langle Y1356\rangle}{\begin{array}{c} \langle Y1235\rangle\langle Y1236\rangle\langle Y1345\rangle\langle Y1347\rangle\langle Y1367\rangle\langle Y1567\rangle\langle Y3456\rangle\langle Y4567\rangle \end{array}},\\
\label{eq:forms2}
\omega_4^{--}&{=}\frac{-\langle Y(12345){\cap}(12367)\rangle\langle Y(12367){\cap}(34567)\rangle^3}{\begin{array}{c}\langle Y1235\rangle\langle Y1237\rangle\langle Y1267\rangle\langle Y1367\rangle\langle Y2367\rangle\langle Y3456\rangle\\\langle Y3467\rangle\langle Y4567\rangle\langle Y123(45){\cap}(Y367)\rangle\end{array}}.
\end{align}
The space corresponding to the form $\Omega^{--}$ has both $\langle Y1235\rangle{=}0$ and $\langle Y1236\rangle{=}0$ as co-dimension one boundaries. In fact, this sign flip region is significantly more complicated geometrically as it has a total of thirteen co-dimension one boundaries (or, equivalently, thirteen poles in the canonical form). Numerically evaluating the sum 
\begin{equation}
\Omega_7^{\text{N}^2\text{MHV}}=\sum_{I{=}(-+,+-,--)}\sum_j\omega^I_j
\end{equation}
we reproduce the BCFW representation (\ref{eq:bcfwresult}), thus verifying the equivalence of the sign flip definition of the space with the original $C\cdot Z$ definition.

The results (\ref{eq:ex7})-(\ref{eq:forms2}) do not immediately suggest an obvious generalization to all multiplicity. However, the collection of forms for the sign pattern $\Omega^{+-}$ is an artifact of our choice of triangulation, and does not make manifest the fact that the $(-+)$ and $(+-)$ regions are equally complicated geometrically. To find a simpler representation of $\Omega^{+-}$ which makes this manifest, we first rewrite our result for the $(-+)$ sign pattern as
\begin{align}
\label{eq:10pattern}
\Omega^{-+}=&\sum_{i=7,1}\frac{(-1)\langle123456\rangle\langle i{-}1ii{+}1345\rangle^3}{\langle Y1236\rangle\langle Y345i{-}1\rangle\langle Y345i{+}1\rangle\langle Y34ii{+}1\rangle\langle Y45i{-}1i\rangle\langle Y5i{-}1ii{+}1\rangle}\nonumber\\ &+\sum_{i=1,3}\frac{\langle123456\rangle\langle Y(12345){\cap}(ii{+}1567)\rangle^3}{\begin{array}{c}\langle Y1236\rangle\langle Y1234\rangle\langle 2345\rangle\langle Y12(34){\cap}(Y567)5\rangle\\\langle Y1(ii{+}1){\cap}(Y567)45\rangle\langle Yii{+}156\rangle\langle Yii{+}157\rangle\end{array}}.
\end{align}
Here, we observe the privileged r\^{o}le of the bracket $\langle Y1236\rangle$ which labels the position of the positive bracket for this pattern. This suggests a natural conjecture for the $(+-)$ sign pattern form using the bracket $\langle Y1235\rangle$:
\begin{align}
\label{eq:01pattern}
\Omega^{+-}=&\sum_{i=2,3}\frac{\langle123567\rangle\langle i{-}1ii{+}1567\rangle^3}{\langle Y1235\rangle\langle Y567i{-}1\rangle\langle Y567i{+}1\rangle\langle Y56ii{+}1\rangle\langle Y67i{-}1i\rangle\langle Y7i{-}1ii{+}1\rangle}\nonumber\\ &+\sum_{i=3,5}\frac{(-1)\langle123567\rangle\langle Y(34567){\cap}(ii{+}1712)\rangle^3}{\begin{array}{c}\langle Y1235\rangle\langle Y3456\rangle\langle Y4567\rangle\langle Y34(56){\cap}(Y712)7\rangle\\\langle Y3(ii{+}1){\cap}(Y712)67\rangle\langle Yii{+}171\rangle\langle Yii{+}172\rangle\end{array}}.
\end{align}
Direct numerical comparison to the representation $\sum_{i=1}^8\omega_i^{+-}$ obtained by brute force establishes the correctness of this ansatz. For the all-minus sign pattern corresponding to $\Omega^{--}$, the set of objects used for the $(+-)$ and $(-+)$ sign patterns is not sufficient. The first step in deducing a compact representation of this form is in factoring the cubic pole in $\omega^{--}_1$ and $\omega_2^{--}$ as 
\begin{align}
 & [2]\langle Y1245\rangle\langle Y123(45){\cap}(Y367)\rangle{+}\langle Y1345\rangle\langle Y(12(45){\cap}(Y123)67){\cap}(34567)\rangle \nonumber\\
 &=\langle Y(12345){\cap}(34567)\rangle\langle Y123(45){\cap}(Y671)\rangle, 
\end{align}
which cancels one factor in the numerator of each term, after which the sum of terms can be compactly written in terms of some simple objects which generalize to the all-multiplicity case smoothly. The schematic form of the result is 
\begin{align}
\Omega^{--}=\sum_{\substack{ijk\ell\\ m=5,6}}\left(\mathcal{O}^{(1)}_{ijk\ell;m}+\mathcal{O}_{ijk\ell}+\mathcal{O}_{ij}\right),
\end{align}
where
\begin{align}
\label{eq:mmObj}
\mathcal{O}^{(1)}_{ijk\ell;m}=&\frac{\langle Y(123mi){\cap}(j{-}1jj{+}1kk{+}1\rangle\langle Y(j{-}1jj{+}1kk{+}1){\cap}(i{-}1ii{+}1\ell\ell{+}1)\rangle^3}{\begin{array}{c}\langle Y123m\rangle\langle Yj{-}1jj{+}1k{+}1\rangle\langle Yj{-}1jkk{+}1\rangle\langle Yj{-}1j{+}1kk{+}1\rangle\langle Yjj{+}1kk{+}1\rangle\\\langle Yi{-}1ii{+}1\ell\rangle  \langle Yii{+}1\ell\ell{+}1\rangle\langle Yi{-}1i\ell\ell{+}1\rangle \langle Yj{-}1jj{+}1(k\ell{+}1){\cap}(Yi{-}1ii{+}1)\rangle \end{array}},\nonumber\\
\mathcal{O}_{ijk\ell}=&\frac{\langle Y(i{-}1ii{+}1jj{+}1){\cap}(k{-}1kk{+}1\ell\ell{+}1)\rangle^4}{\begin{array}{c}\langle Yi{-}1ii{+}1j\rangle\langle Yi{-}1ii{+}1j{+}1\rangle\langle Yii{+}1jj{+}1\rangle\langle Yk{-}1kk{+}1\ell\rangle\langle Yk{-}1kk{+}1\ell{+}1\rangle\\\langle Yk{-}1k\ell\ell{+}1\rangle\langle Yi{-}1ii{+}1(\ell\ell{+}1){\cap}(Yk{+}1jj{+}1)\rangle\langle Yk{-}1kk{+}1(jj{+}1){\cap}(Yi{-}1\ell\ell{+}1)\rangle\end{array}},\nonumber\\
\mathcal{O}_{ij}=&\frac{\langle i{-}1ii{+}1j{-}1jj{+}1\rangle^3\langle Y(1235i){\cap}(1236j)\rangle\langle Yi{-}1i{+}1j{-}1j{+}1\rangle}{\begin{array}{c}\langle Y1235\rangle\langle Y1236\rangle\langle Yi{-}1ii{+}1j{-}1\rangle\langle Yii{+}1j{-}1j\rangle\langle Yi{+}1j{-}1jj{+}1\rangle\\\langle Yj{-}1jj{+}1i{-}1\rangle\langle Yi{-}1ijj{+}1\rangle\langle i{-}1ii{+}1j{+}1\rangle\end{array}},
\end{align}
and at seven point $m=5,6$ and $m\neq1,2,3,n$ more generally. 
\subsubsection{All multiplicity generalization} \label{sec:n2mhvAlln}
A compact formula for the $n$-point $\text{N}^2$MHV tree-level amplitude can be generated using the CSW recursion relations \cite{Cachazo:2004kj,Bena:2004}, which was reformulated in momentum twistor space using MHV diagrams in \cite{Bullimore:2010pj}. In terms of the reference twistor $Z_{\star}$ the super-amplitude is a (cyclic) sum of products of shifted $R$-invariants, \begin{equation}\label{eq:csw1} A_n^{\text{N}^2\text{MHV}}=\sum_{i<j\leq k<\ell\leq i} [\star,\hat{i},i{+}1,j,j{+}1]\times[\star,\hat{k},k{+}1,\ell,\ell{+}1],\end{equation}
where in ordinary momentum twistor space the $R$-invariant $[abcde]$ is given by 
\begin{equation}\label{eq:R} [abcde]=\frac{\delta^{0\vert4}\left(\langle abcd\rangle\eta_e+\mathrm{cyclic}\right)}{\langle abcd\rangle\langle bcde\rangle\langle cdea\rangle\langle deab\rangle\langle eabc\rangle}.\end{equation} 
In (\ref{eq:csw1}) the shifted twistors $\hat{Z}_i$ and $\hat{Z}_k$ are defined in terms of the intersections of lines and planes: \begin{equation} \label{eq:shift}\begin{split} \hat{i}&=\left\{\begin{array}{lr} (ii{+}1)\cap(\star kk{+}1) & i=\ell \\ 
						    i & \text{otherwise}\end{array}\right.,\\
\hat{k}&=\left\{\begin{array}{lr} (kk{+}1)\cap(\star ii{+}1) & j=k \\ 
						   k & \text{otherwise}\end{array}\right. .	    \end{split}\end{equation}
In the $Y$-space of the amplituhedron, the product of $R$-invariants entangles the numerators and the explicit formula becomes: 
 \begin{equation}\label{eq:csw} \Omega_n^{\text{N}^2\text{MHV}}=\sum_{i<j\leq k<\ell\leq i} \frac{\langle Y(\star ii{+}1jj{+}1){\cap}(\star kk{+}1\ell\ell{+}1)\rangle^4}{\begin{array}{c}\langle Y{\star}ii{+}1j\rangle\langle Y{\star} ii{+}1j{+}1\rangle\langle Y{\star} ijj{+}1\rangle\langle Y{\star} i{+}1jj{+}1\rangle\langle Yii{+}1jj{+}1\rangle \\ \langle Y{\star}kk{+}1\ell\rangle\langle Y{\star} kk{+}1\ell{+}1\rangle\langle Y{\star}k\ell\ell{+}1\rangle\langle Y{\star}k{+}1\ell\ell{+}1\rangle\langle Ykk{+}1\ell\ell{+}1\rangle\end{array}}, \end{equation}
  where for the boundary case $k=j$ the term is modified to 
  \begin{equation}
  \label{eq:boundary1m4k2} 
  \frac{\langle Y(\star ii{+}1jj{+}1){\cap}(\star jj{+}1\ell\ell{+}1)\rangle^4}{\begin{array}{c}\langle Y{\star}ii{+}1j\rangle\langle Y{\star}ijj{+}1\rangle\langle Y{\star}i{+}1jj{+}1\rangle\langle Yii{+}1jj{+}1\rangle\langle Y{\star}jj{+}1\ell\rangle\\\langle Y{\star}jj{+}1\ell{+}1\rangle \langle Y{\star}j{+}1\ell\ell{+}1\rangle\langle Yjj{+}1\ell\ell{+}1\rangle\langle Y{\star}ii{+}1(jj{+}1){\cap}(Y{\star}\ell\ell{+}1)\rangle\end{array}}, 
  \end{equation} 
  and for $\ell=i$ the modification is
  \begin{equation} 
  \label{eq:boundary1m4k22}
  \frac{\langle Y(\star ii{+}1jj{+}1){\cap}(\star kk{+}1ii{+}1\rangle^4}{\begin{array}{c} \langle Y{\star} ii{+}1j\rangle\langle Y{\star} ii{+}1j{+}1\rangle\langle Y{\star} ijj{+}1\rangle\langle Yii{+}1jj{+}1\rangle\langle Y{\star} kk{+}1i{+}1\rangle\\\langle Y{\star} kii{+}1\rangle  \langle Y{\star} k{+}1ii{+}1\rangle\langle Ykk{+}1ii{+}1\rangle\langle Y{\star}jj{+}1(ii{+}1){\cap}(Y{\star}kk{+}1)\rangle \end{array}}. \end{equation} 
  
From the sign flip perspective, the $n$-point amplituhedron is a collection of spaces labelled by the sequence $\{\langle Y123i\rangle\}_{i=5,\ldots,n{-}1}$: 
\begin{align}
\mathcal{A}^{\text{N}^2\text{MHV}}_n{=}\{&(-,+,\ldots,+),\ldots,(+,+,\ldots,-),(-,-,+,\ldots,+),\nonumber\\ &\ldots,(+,\ldots,-,-),\ldots,(-,\ldots,-)\}.
\end{align}
The canonical forms for single-plus or single-minus sign patterns $(-,+,\ldots,+)$ are given by straightforward extensions of the seven point cases (\ref{eq:10pattern}) and (\ref{eq:01pattern}), while the multiple minus-sign pattern forms are expressible in terms of the objects (\ref{eq:mmObj}) where the important brackets are those labelling the flips $+\leftrightarrow-$. In fact, the collection of objects defined in the previous section for the seven point sign patterns is sufficiently general to match the amplitude, a fact we verified by comparison to the CSW expansion (\ref{eq:csw}) up to $n=12$. Our result for the $n$-point tree-level canonical form can be written as 
\begin{equation}
\Omega_n^{\text{N}^2\text{MHV}}=\sum_{ijk\ell m}\left(\mathcal{O}_{ij;k}+\mathcal{O}^{(1)}_{ijk\ell;m}+\mathcal{O}^{(2)}_{ijk\ell;m}+\mathcal{O}_{ijk\ell}+\mathcal{O}_{ij;k\ell}+\mathcal{O}_{ij}\right),
\end{equation}
where the various forms needed to match the amplitude (which are not already defined above) are
\begin{align}
\mathcal{O}_{ij;k}{=}&\frac{\langle123jj{+}1\rangle\langle i{-}1ii{+}1j{-}1jj{+}1\rangle^3}{\langle Y123k\rangle\langle Yj{-}1jj{+}1i{-}1\rangle\langle Yj{-}1jj{+}1i{+}1\rangle\langle Yj{-}1jii{+}1\rangle\langle Yjj{+}1i{-}1i\rangle\langle Yj{+}1i{-}1ii{+}1\rangle},\nonumber\\
\mathcal{O}^{(2)}_{ijk\ell;m}{=}&\frac{\langle 123mii{+}1\rangle\langle Y(j{-}1jj{+}1ii{+}1){\cap}(kk{+}1\ell{-}1\ell\ell{+}1)\rangle^3}{\begin{array}{c}\langle Y123m\rangle\langle Yj{-}1jj{+}1i\rangle\langle Yjj{+}1ii{+}1\rangle\langle Yj{-}1j(j{+}1i){\cap}(Y\ell{-}1\ell\ell{+}1)j{+}1\rangle \\ \langle Yj{-}1(kk{+}1){\cap}(Y\ell{-}1\ell\ell{+}1)ii{+}1\rangle\langle Ykk{+}1i{+}1\ell\rangle\langle Ykk{+}1\ell{-}1\ell{+}1\rangle\end{array}},\nonumber\\
\mathcal{O}_{ij;k\ell}=&\frac{\langle i{-}1ii{+}1j{-}1jj{+}1\rangle^3\langle Y(123ki){\cap}(123\ell j)\rangle\langle Yi{-}1i{+}1j{-}1j{+}1\rangle}{\begin{array}{c}\langle Y123k\rangle\langle Y123\ell\rangle\langle Yi{-}1ii{+}1j{-}1\rangle\langle Yii{+}1j{-}1j\rangle\langle Yi{+}1j{-}1jj{+}1\rangle\\\langle Yj{-}1jj{+}1i{-}1\rangle\langle Yi{-}1ijj{+}1\rangle\langle i{-}1ii{+}1j{+}1\rangle\end{array}}.
\end{align}

\section{$6{\times}2$ representation of the one-loop NMHV amplituhedron}
\label{sec:4}

\subsection{NMHV one-loop as a product of $m{=}2$ amplituhedra}
The sign flip definition of the one-loop NMHV amplituhedron is 
\begin{equation}
\begin{split}
&\langle(YAB)ii{+}1\rangle>0,\langle Yii{+}1jj{+}1\rangle>0,\\
&\{\langle(YAB)1i\rangle\}_{i=2,\ldots,n} \quad \text{has three sign flips},\\
&\{\langle Y123i\rangle\}_{i=4,\ldots,n}\quad \text{has one sign flip}.
\end{split}
\end{equation}
From this definition, we can see the one-loop NMHV amplituhedron naturally factors into a product of two $m=2$ amplituhedra; namely, the $m=2,k=3$ amplituhedron in $(YAB)$ and the polygon which is the intersection of the plane-$(YAB)$ and the $k=1$ tree amplituhedron i.e., the convex hull of the external data.  Since this intersection is simply a polygon, the only remaining constraint is that the point $Y$ on the plane $(YAB)$ lie inside the polygon. This implies the canonical form of the one-loop NMHV space can be expressed as the product of a six-form and a two-form, where the six-form is the canonical form for the plane $(YAB)$, and the two-form is in the point contained inside the intersecting polygon. The important point is that in this representation, there is effectively no difference between the tree and loop-level variables. To fix notation, we will write the $(YAB)$ plane as $\mathcal{Y}=(Y_1Y_2Y_3)=(YAB)$, the span of three points, and the point on the intersecting polygon as $y$, which has two degrees of freedom.

This $6\times2$ representation is to be contrasted with the usual $4\times4$ representation of the NMHV one-loop canonical form. From the original $Y=C\cdot Z$ definition, the canonical form is written in terms of the four-form in the point $Y$ and the four-form in the line $(AB)$. This corresponds to the usual BCFW representation of the loop integrand, which is organized as ($R$-invariant)$\times$(loop form in $(AB)$). As we shall see, triangulating the space using the $6\times2$ picture yields a new representation, distinct from BCFW or any other recursion, which suggests different organizing principles for further calculations.

 Let us begin by identifying the vertices of the intersecting polygon. Consider the intersection of the three-plane $\mathcal{Y}$ and the four-dimensional cyclic polytope with vertices ${Z_i}$. The boundaries of this polygon are determined by the intersection of the $\mathcal{Y}$-plane and the facets of the cyclic polytope, $(ii{+}1 jj{+}1)$. Each vertex is the intersection of the $\mathcal{Y}$-plane and a two-plane labelled by three indices shared by two boundaries of the cyclic polytope. For example, the plane defined by the two boundaries $(ii{+}1jj{+}1),(ii{+}1j{+}1j{+}2)$ is $(ii{+}1j{+}1)$. Explicitly, a boundary of this polytope $(ii{+}1jj{+}1)$ intersects the $\mathcal{Y}$-plane in a line
 \begin{equation}
\mathcal{Y}{\cap}(ii{+}1jj{+}1){=}(ii{+}1)\langle \mathcal{Y}jj{+}1\rangle+(i{+}1j)\langle \mathcal{Y}j{+}1i\rangle+(jj{+}1)\langle \mathcal{Y}ii{+}1\rangle+(j{+}1i)\langle \mathcal{Y}i{+}1j\rangle.
\end{equation}
Similarly, the plane $(ii{+}1j)$ intersects $\mathcal{Y}$ in a point
\begin{equation}
\mathcal{Y}\cap (ii{+}1j)=Z_i\langle \mathcal{Y}i{+}1j\rangle+Z_{i+1}\langle \mathcal{Y}ji\rangle+Z_j\langle \mathcal{Y}ii{+}1\rangle.
\end{equation}
This point is in the interior of this polytope if all of these coefficients are positive,
\begin{equation}
\label{eq:vertexcondition}
\langle \mathcal{Y}ii{+}1\rangle,\langle \mathcal{Y}i{+}1j\rangle,\langle \mathcal{Y}ji\rangle>0.
\end{equation}
To summarize, the vertices of the intersecting polygon are labelled by triplets $(a,b,c)$ which satisfy \eqref{eq:vertexcondition}. The analogous case in higher dimensions is discussed in \cite{Arkani-Hamed:2018ign}.

Once we obtain the vertices of the intersecting polygon, the corresponding canonical form for the point $y$ inside is straightforward to compute. For example, the logarithmic two-form in $y$ of the triangle whose vertices are $\{\hat{i},\hat{j},\hat{k}\}=\{(i_1i_2i_3), (j_1j_2j_3), (k_1k_2k_3)\}$ is
\begin{equation}
\label{eq:triangleform}
\Omega^{(2)}(\hat{i},\hat{j},\hat{k})=\frac{\langle y\mathrm{d}^2y\rangle\langle\hat{i}\hat{j}\hat{k}\rangle^2 }{\langle y\hat{i}\hat{j}\rangle\langle y\hat{j}\hat{k}\rangle\langle y\hat{k}\hat{i}\rangle}.
\end{equation}
Once the forms in $\mathcal{Y}$ and $y$ are known, it is trivial to rewrite this 6$\times$2 representation in the original $(YAB)$ space. First, note that the line $(\hat{i}\hat{j})$ on the plane $\mathcal{Y}$ is the intersection of two boundaries of the cyclic polytope $(i_1i_2i_3)\cap(j_1j_2j_3)$. Similarly, the vertex $(\hat{i}\hat{j}\hat{k})$ is the intersection of three planes $(i_1i_2i_3)\cap(j_1j_2j_3)\cap(k_1k_2k_3)$. From this, the explicit map between the $(\mathcal{Y},y)$ and $(Y,YAB)$ variables is given by
\begin{align}
\label{eq:translation}
\langle \mathcal{Y} ij \rangle =&\langle YAB ij \rangle,\\
\langle y\hat{i}\hat{j}\rangle=&\langle YAB(i_1i_2i_3)\cap(j_1j_2j_3)\rangle\nonumber\\
=&\langle YABi_1i_2\rangle\langle Yi_3j_1j_2j_3\rangle+\langle YABi_2i_3\rangle\langle Yi_1j_1j_2j_3\rangle+\langle YABi_3i_1\rangle\langle Yi_2j_1j_2j_3\rangle,\\
\langle \hat{i}\hat{j}\hat{k}\rangle=&\langle (YAB)\cap(i_1i_2i_3)\cap(j_1j_2j_3)\cap(k_1k_2k_3)\rangle \nonumber\\
=&
\begin{vmatrix} 
 \langle YAi_1i_2i_3\rangle&\langle YAj_1j_2j_3\rangle&\langle YAk_1k_2k_3\rangle
\\ \langle ABi_1i_2i_3\rangle&\langle ABj_1j_2j_3\rangle&\langle ABk_1k_2k_3\rangle\\ \langle BYi_1i_2i_3\rangle&\langle BYj_1j_2j_3\rangle&\langle BYk_1k_2k_3\rangle
\end{vmatrix},
\end{align}
while the measure is modified as
\begin{equation}
\langle \mathcal{Y}\mathrm{d}^2Y_1 \rangle\langle \mathcal{Y}\mathrm{d}^2Y_2 \rangle\langle \mathcal{Y}\mathrm{d}^2Y_3 \rangle\langle y\mathrm{d}^2y \rangle=\langle Y\mathrm{d}^4Y \rangle\langle YAB\mathrm{d}^2A \rangle\langle YAB\mathrm{d}^2B \rangle.
\end{equation}
However, in all subsequent expressions of this section we suppress all such measure factors, which should be clear from context.

This procedure can be straightforwardly generalized to a product representation for the one-loop $\text{N}^k$MHV amplituhedron $\mathcal{A}^{1\text{-loop}}_{n,k}$. From the sign flip definition, it follows that the $\mathcal{A}^{1\text{-loop}}_{n,k}$ space can be constructed from the $m=2,k{+}2$ tree-level space and the intersection of the $m=2,k$ space with the $m=4$ $\text{N}^k$MHV tree amplituhedron. The canonical form of $\mathcal{A}^{1\text{-loop}}_{n,k}$ factorizes to a $2(k{+}2)\times2k$ form, yielding a ``$2(k{+}2)\times2k$'' representation of the geometry.
\subsection{Five point case}
\begin{figure}[t]
\begin{center}
\includegraphics[clip,width=6.0cm]{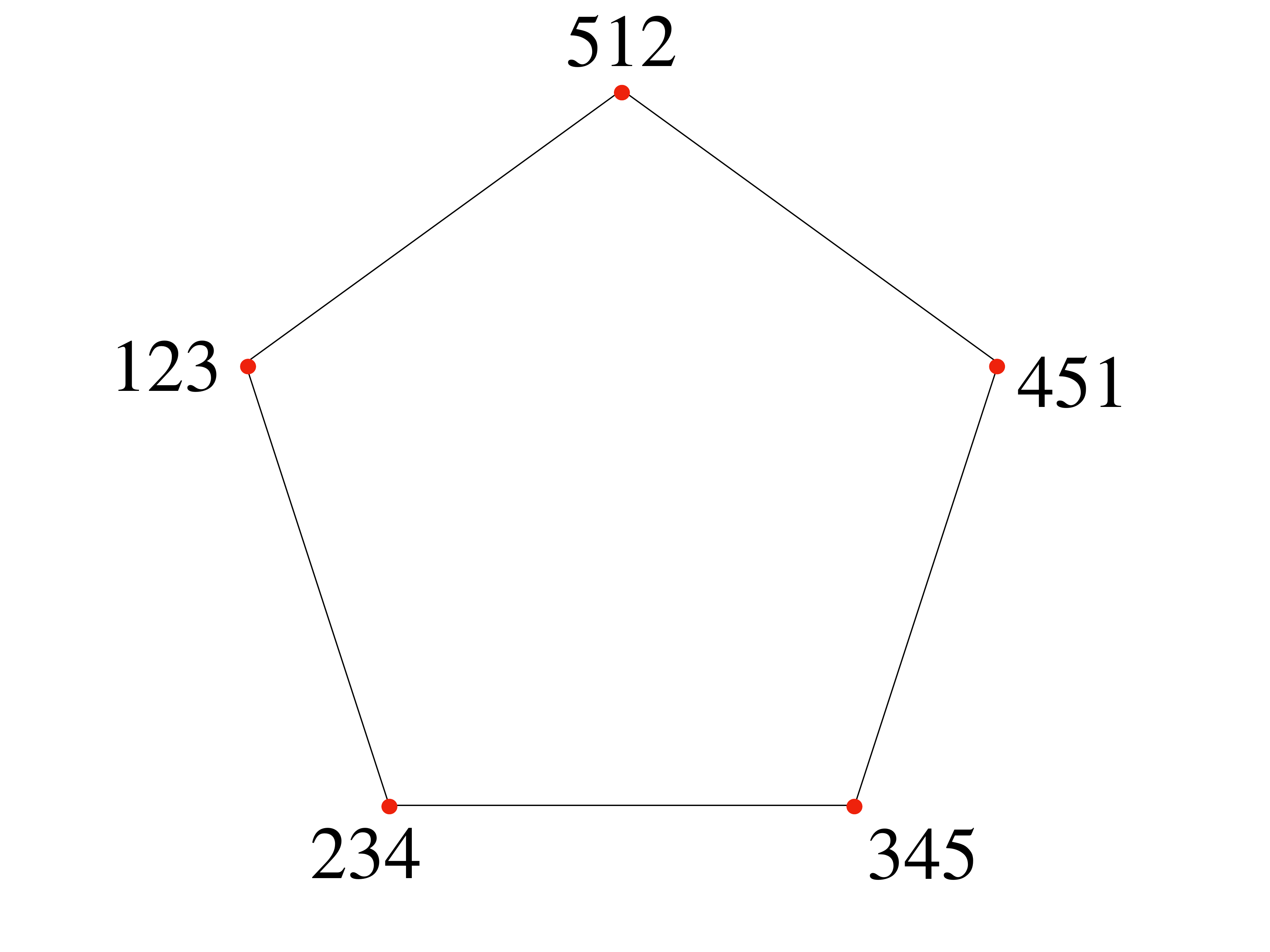}
 \caption{5-pt intersecting pentagon}
 \label{fig:pentagon5pt}
\end{center}
  \end{figure}

In this section, we construct the 6$\times$2 representation of the one-loop five-point NMHV amplituhedron explicitly. In this simplest case, the $m=2,k=3$ amplituhedron is simply the $G_+(3,5)$ positive Grassmannian. The intersecting pentagon has edges associated with the boundaries of the cyclic polytope
\begin{equation}
 (1234),(2345),(3451),(4512),(5123).
 \end{equation}
The triplets defining the possible vertices of the pentagon are
\begin{equation}
(123),(234),(345),(451),(512),(124),(134),(135),(235),(245),(135).
\end{equation}
However, from the previous section a triplet $(a,b,c)$ is a vertex of the intersecting pentagon if the condition \eqref{eq:vertexcondition} is satisfied. From this, we can see that only cyclic triplets $(123),(234),(345),(451),(512)$ label the vertices of the pentagon, which is shown in Figure~\ref{fig:pentagon5pt}.
For notational convenience we label these vertices as $\hat{i}:=(i{-}1ii{+}1)$. From these considerations, it is clear the intersecting pentagon is the $m=2,k=1,n=5$ amplituhedron where the vertices are $(\hat{1},\hat{2},\hat{3},\hat{4},\hat{5})$,
and the 6$\times$2 representation of the five-point one-loop NMHV canonical form is 
\begin{equation}
\label{eq:5pt62rep}
\mathcal{A}^{\text{1-loop}}_{5,1}=\mathcal{A}^{m=2,\text{tree}}_{5,3}(1,\ldots,5)\times \mathcal{A}^{m=2,\text{tree}}_{5,1}(\hat{1},\ldots,\hat{5}).
\end{equation}
This corresponds directly to the representation obtained from the momentum twistor diagrams of \cite{Bai:2015qoa}. From this 6$\times$2 representation, we can see the geometric factor of the measure of the one-loop NMHV amplituhedron is nothing but the intersecting $m=2,k=1$ tree amplituhedron \cite{Bai:2015qoa}. 

From (\ref{eq:5pt62rep}) it is trivial to compute the $6\times2$ canonical form; first, the six-form is simply the top form on $G_+(3,5)$,
\begin{equation}
\Omega^{(6)}(\mathcal{Y})=\frac{\langle 12345 \rangle^2}{\langle \mathcal{Y}12 \rangle\langle \mathcal{Y}23 \rangle\langle \mathcal{Y}34 \rangle\langle \mathcal{Y}45 \rangle\langle \mathcal{Y}51 \rangle}.
\end{equation}
To obtain the canonical form of the intersecting pentagon, we need to triangulate it. This can be done by, for example, using the lines $\hat{1}\hat{3}$ and $\hat{1}\hat{4}$ in Figure~\ref{fig:pentagon5pt}, which gives
\begin{equation}
\Omega^{(2)}(y)=\frac{\langle \hat{1}\hat{2}\hat{3}\rangle^2}{\langle y\hat{1}\hat{2}\rangle\langle y\hat{2}\hat{3}\rangle\langle y\hat{1}\hat{3}\rangle}+\frac{\langle \hat{1}\hat{3}\hat{4}\rangle^2}{\langle y\hat{1}\hat{3}\rangle\langle y\hat{3}\hat{4}\rangle\langle y\hat{4}\hat{5}\rangle}+\frac{\langle \hat{1}\hat{4}\hat{5}\rangle^2}{\langle y\hat{1}\hat{4}\rangle\langle y\hat{4}\hat{5}\rangle\langle y\hat{1}\hat{5}\rangle}.
\end{equation}
The full $6\times2$ form is given by the product
\begin{equation}
\label{eq:62formmomentum}
\Omega^{5,1,1}(\mathcal{Y},y)=\Omega^{(6)}(\mathcal{Y})\times\Omega^{(2)}(y).
\end{equation}
We can transform back into $(YAB)$ space by using the map \eqref{eq:translation}, and the result is
 \begin{equation}
 \begin{split}
\Omega^{5,1,1}(Y,(YAB))=&\frac{\langle 12345 \rangle^2}{\langle YAB12 \rangle\langle YAB23 \rangle\langle YAB34 \rangle\langle YAB45 \rangle\langle YAB51 \rangle}\\
&\times\Bigg(\frac{\langle YAB12 \rangle\langle YAB23 \rangle\langle 12345 \rangle^2}{\langle Y1235 \rangle\langle Y1234 \rangle\langle YAB13 \rangle\langle YAB(125){\cap}(234) \rangle}\\
&+\frac{\langle YAB45 \rangle\langle YAB15 \rangle\langle 12345 \rangle^2}{\langle Y3451 \rangle\langle Y4512 \rangle\langle YAB14 \rangle\langle YAB(512){\cap}(345) \rangle}\\
&+\frac{\langle YAB34 \rangle\langle YAB25 \rangle^2\langle 12345 \rangle^2}{\langle Y2345 \rangle\langle Y3451 \rangle\langle YAB13 \rangle\langle YAB45 \rangle\langle YAB(125){\cap}(234) \rangle}\Bigg).
 \end{split}
\end{equation}
Of course, we can triangulate the pentagon in another way. For example, using the lines $(\hat{5}\hat{2}),\ (\hat{5}\hat{3})$ to triangulate the space, we obtain
\begin{equation}
\Omega^{(2)}(y)=\frac{\langle \hat{5}\hat{1}\hat{2}\rangle^2}{\langle y\hat{5}\hat{1}\rangle\langle y\hat{1}\hat{2}\rangle\langle y\hat{5}\hat{2}\rangle}+\frac{\langle \hat{5}\hat{2}\hat{3}\rangle^2}{\langle y\hat{5}\hat{2}\rangle\langle y\hat{2}\hat{3}\rangle\langle y\hat{5}\hat{3}\rangle}+\frac{\langle \hat{5}\hat{3}\hat{4}\rangle^2}{\langle y\hat{5}\hat{3}\rangle\langle y\hat{3}\hat{4}\rangle\langle y\hat{5}\hat{4}\rangle},
\end{equation}
which when combined with $\Omega^{(6)}(\mathcal{Y})$ and rewritten in the $(YAB)$ space, gives exactly the BCFW representation of the five-point integrand, 
 \begin{align}
\Omega^{5,1,1}(Y,(YAB))=&\frac{\langle 12345 \rangle^4}{\langle Y1245 \rangle\langle Y1235 \rangle\langle YAB23 \rangle\langle YAB34 \rangle\langle YAB45 \rangle\langle YAB(145){\cap}(123) \rangle}\nonumber\\
&+\frac{\langle 12345 \rangle^4}{\langle Y1345 \rangle\langle Y2345 \rangle\langle YAB12 \rangle\langle YAB23 \rangle\langle YAB15 \rangle\langle YAB(145)\cap(234) \rangle}\nonumber\\
&+\frac{\langle 12345 \rangle^4\langle YAB14 \rangle^2}{\begin{array}{c}\langle Y1234 \rangle\langle YAB12 \rangle\langle YAB34 \rangle\langle YAB45 \rangle\langle YAB15 \rangle\\\langle YAB(145){\cap}(123) \rangle\langle YAB(145){\cap}(234) \rangle\end{array}}.
 \end{align}
However, the fact that the BCFW triangulation can be interpreted as one of the
triangulations of the intersecting pentagon holds only for the five point case. At higher points there seems to be no triangulation of the intersecting polygon which corresponds to the BCFW representation of the integrand. 
\subsection{Six point case}\label{sec:6ptNMHV1loop}
\begin{figure}[t]
\begin{center}
 \includegraphics[clip,width=14.0cm]{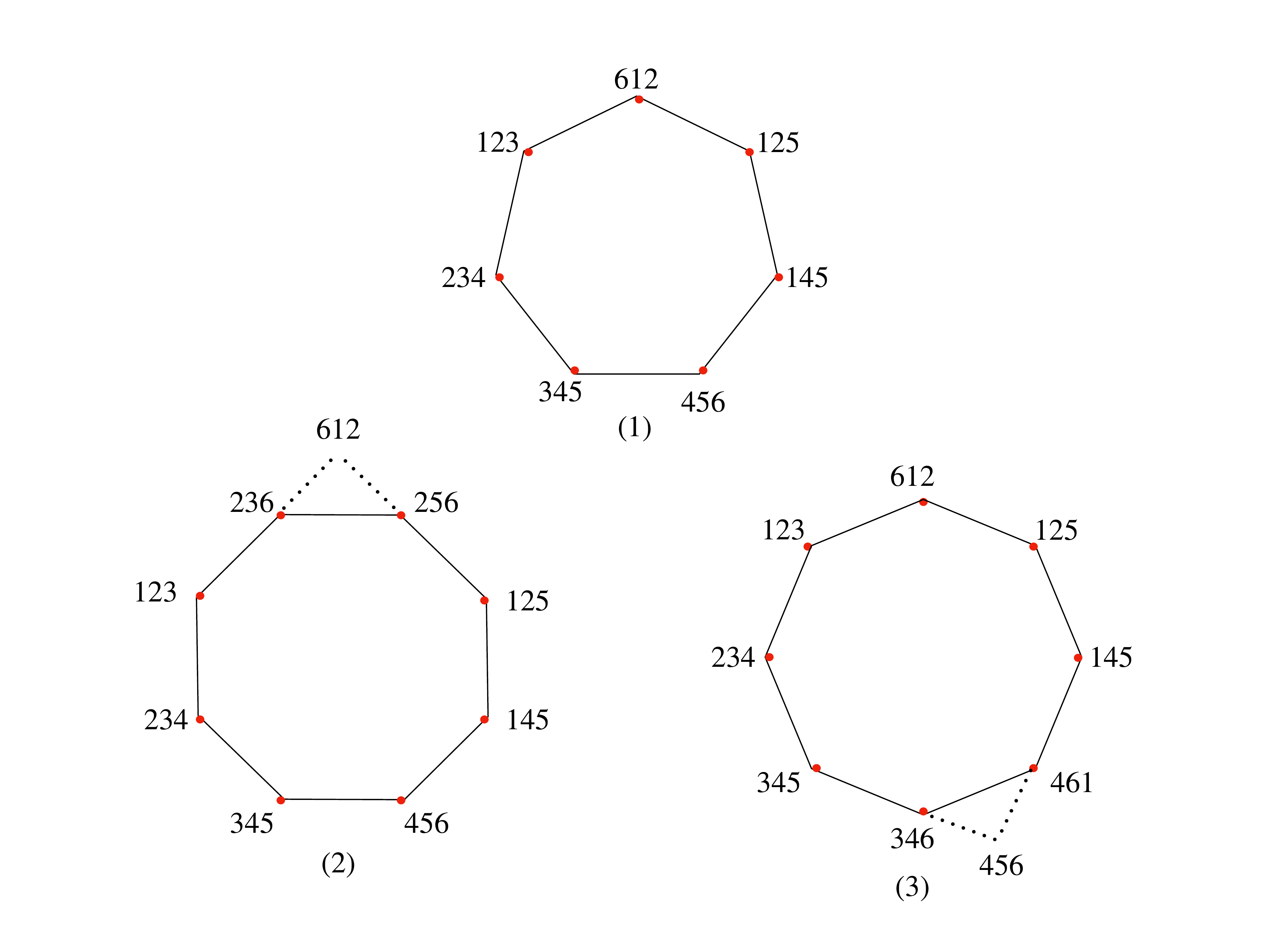}
 \end{center}
 \caption{Polygons for each six-point cell in $A_{234}$.}
 \label{fig:234polygon}
 \end{figure}

At six points, the shape of the intersecting hexagon depends on the positivity conditions involving $\mathcal{Y}$, and the triangulation is nontrivial both in $\mathcal{Y}$ and $y$. First, the list of triplets generated by the intersection of two facets of the cyclic polytope is
\begin{equation}
 \begin{split}
&(123),(124),(125),(134),(234),(235),(236),(345),(346)\nonumber\\
&(456),(561),(612),(245),(356),(256),(461),(136),(145).
 \end{split}
\end{equation}
From the sign flip definition, the $\mathcal{Y}$-space amplituhedron is decomposed into four cells as summarized in the following table:
\begin{center}
\begin{tabular}{| c | c | c | c | c | c |}
\hline
  &$\langle \mathcal{Y}12\rangle$&$\langle \mathcal{Y}13\rangle$&$\langle \mathcal{Y}14\rangle$&$\langle \mathcal{Y}15\rangle$&$\langle \mathcal{Y}16\rangle$\\\hline
$\mathcal{A}_{234}$&+&$-$&+&$-$&$-$\\\cline{1-6}
$\mathcal{A}_{235}$&+&$-$&+&+&$-$\\\cline{1-6}
$\mathcal{A}_{245}$&+&$-$&$-$&+&$-$\\\cline{1-6}
$\mathcal{A}_{345}$&+&+&$-$&+&$-$\\\cline{1-6}
\end{tabular}
\end{center}
Here we label individual sign patterns by the places where the sign flips occur. For example, the $\mathcal{A}_{234}$ cell indicates the sign flips are between the $\langle \mathcal{Y}12\rangle,\langle \mathcal{Y}13\rangle$ and $\langle \mathcal{Y}14\rangle$ positions in the sequence. Let us consider this particular cell in more detail. From the signs of the brackets $\langle \mathcal{Y}ii{+}1\rangle,\langle \mathcal{Y}1i \rangle$, we can see that although $(123),(125),(234),(345),(145)$ can be vertices of the polygon, whether the other planes $(236),(346),(456),(612),(256),(461)$ can be vertices depends on the signs of the additional brackets $\langle \mathcal{Y}26\rangle,\ \langle \mathcal{Y}36\rangle$ and $ \langle \mathcal{Y}46\rangle$. The possible sign patterns of these brackets dictates the shape of the intersecting polygon, and the different cases are given in the following table:
\begin{center}
\begin{tabular}{| c | c | c | c | c |}
\hline
 $\langle \mathcal{Y}26\rangle$&$\langle \mathcal{Y}36\rangle$&$\langle \mathcal{Y}46\rangle$&vertices&pentagon\\\hline
+&+&$-$&$(612),(456)$&\multirow{2}{*}{$(1)$}\\\cline{1-4}
+&$-$&$-$&$(612),(456)$&\\\cline{1-5}
$-$&+&$-$&$(236),(256),(456)$&$(2)$\\\cline{1-5}
+&$-$&+&$(346),(456),(612)$&$(3)$\\\cline{1-5}
\end{tabular}
\end{center}
For each sign pattern, there is an associated polygon as shown in Figure \ref{fig:234polygon}.
To obtain the canonical form associated to the space $\mathcal{A}_{234}$, we need to further triangulate each intersecting polygon by introducing additional lines, just as at five points. Repeating this process for the three remaining cells $\mathcal{A}_{235},\mathcal{A}_{245},\mathcal{A}_{345}$ yields the full result; we leave the details of the calculation to appendix \ref{sec:appendix1}, and write here only the final result for the space:
\begin{align}
\label{eq:6pt62rep}
\mathcal{A}^{\text{1-loop}}_{6,1}=&\mathcal{A}^{m=2,\text{tree}}_{6,3}(1,\ldots,6)\times \mathcal{A}^{m=2}_{6,1}(\hat{1}\ldots,\hat{6})\nonumber\\
&{+}\sum_{1\leq i \leq6}\mathcal{A}^{m=2,\text{tree}}_{5,3}(i{+}2,i{+}3,\cdots,i{-}1,i){\times} \mathcal{A}^{(2)}((ii{+}1i+2),(ii{+}2i{+}3),(ii{+}2i{-}1)),
 \end{align}
where $\mathcal{A}^{(2)}(a,b,c)$ is the two-dimensional triangle with vertices $a,b,c$ (whose canonical form is $\Omega^{(2)}(a,b,c)$ which was defined in \eqref{eq:triangleform}). 

To obtain a representation of the canonical form, we need to triangulate the $m=2,k=3$ amplituhedron. Fortunately, we can triangulate all $m=2$ amplituhedra by entangling $k$ copies of the $k=1$ form \cite{Arkani-Hamed:2017vfh}, so the canonical form for arbitrary $k$ can be written as
\begin{equation}
\label{eq:m3nkForm}
\Omega^{m=2}_{n,k}=\sum_{2\leq j_1\leq j_2\leq\cdots\leq j_k\leq n-1}[1,j_1,j_1{+}1;1,j_2,j_2{+}1;\ldots;1,j_k,j_k{+}1],
\end{equation}
where for $k=3$ we have explicitly
\begin{equation}
[i_1,i_2,i_3;\ldots;k_1,k_2,k_3]:=\frac{\langle\mathcal{Y}\cap(i_1i_2i_3)\cap\cdots\cap(k_1k_2k_3)\rangle}{\langle \mathcal{Y}i_1i_2 \rangle\langle \mathcal{Y}i_2i_3 \rangle\langle \mathcal{Y}i_3i_1 \rangle\cdots\langle \mathcal{Y}k_3k_1 \rangle}.
\end{equation}
From this, we can write the canonical form of the 6$\times$2 representation at six points as 
\begin{align}
 \label{eq:6pt}
\Omega^{\text{1-loop}}_{6,1}(\mathcal{Y},y)=&\Omega_{6,3}^{m=2}(\mathcal{Y},(61))\times\sum_{2\leq i\leq 5}\Omega^{(2)}(1,i,i{+}1)\nonumber\\
&{+}\sum_{1\leq i\leq 6}\Omega_{5,3}^{m=2}(i,i{+}2)\times\Omega^{(2)}((ii{+}2i{-}1),(ii{+}2i{+}1),(ii{+}2i{+}3)),
 \end{align}
where we defined
\begin{equation}
\label{eq:triangulationnpt2}
\Omega^{m=2}_{n,k}(a,b):=\sum_{\substack{b{+}1\leq i_1<\cdots<i_k\leq a{-}1}}[b,i_1,i_1{+}1;\ldots;b,i_k,i_k{+}1].
\end{equation}
We can transform into $(YAB)$ space by using the map \eqref{eq:translation}. The explicit expression for the 6$\times$2 representation of the six-point one-loop NMHV amplituhedron canonical form is
\begin{equation}
\begin{split}
\Omega^{\text{1-loop}}_{6,1}(Y,YAB)=&\left(\Omega'_{234}+\Omega'_{235}+\Omega'_{245}+\Omega'_{345}\right)\times\left([123]+[134]+[145]+[156]\right)\\
&+\frac{\langle 12345 \rangle^2\langle 12456 \rangle^2}{\langle YAB 12 \rangle\langle YAB 23 \rangle\langle YAB 34 \rangle\langle YAB 45 \rangle\langle Y 1245 \rangle\langle Y 1256 \rangle\langle Y 4561 \rangle}\\
&+\frac{\langle 13456 \rangle^2\langle 12346 \rangle^2}{\langle YAB 34 \rangle\langle YAB 45 \rangle\langle YAB 56 \rangle\langle YAB 61 \rangle\langle Y 1234 \rangle\langle Y 3461 \rangle\langle Y 2361 \rangle}\\
&+\frac{\langle 12346 \rangle^2\langle 13456 \rangle^2}{\langle YAB 12 \rangle\langle YAB 23 \rangle\langle YAB 34 \rangle\langle YAB 61 \rangle\langle Y 4561 \rangle\langle Y 3461 \rangle\langle Y 3456 \rangle}\\
&+\frac{\langle 12356 \rangle^2\langle 23456 \rangle^2}{\langle YAB 12 \rangle\langle YAB 23 \rangle\langle YAB 56 \rangle\langle YAB 61 \rangle\langle Y 2356 \rangle\langle Y 2345 \rangle\langle Y 3456 \rangle}\\
&+\frac{\langle 12456 \rangle^2\langle 12345 \rangle^2}{\langle YAB 12 \rangle\langle YAB 45 \rangle\langle YAB 56 \rangle\langle YAB 61 \rangle\langle Y 1234 \rangle\langle Y 2345 \rangle\langle Y 1245 \rangle}\\
&+\frac{\langle 23456 \rangle^2\langle 12356 \rangle^2}{\langle YAB 23 \rangle\langle YAB 34 \rangle\langle YAB 45 \rangle\langle YAB 56 \rangle\langle Y 2361 \rangle\langle Y 2356 \rangle\langle Y 1256 \rangle},
 \end{split}
\end{equation}
where
\begin{equation}
 \begin{split}
\Omega'_{ijk}&=\frac{\begin{vmatrix} 
 \langle YA1ii+1\rangle&\langle YA1jj+1\rangle\langle YA1kk+1\rangle
\\ \langle AB1ii+1\rangle&\langle AB1jj+1\rangle\langle AB1kk+1\rangle\\ \langle BY1ii+1\rangle&\langle BY1jj+1\rangle\langle BY1kk+1\rangle
\end{vmatrix}^2}{\begin{array}{c}\langle YAB1i\rangle\langle YAB1i{+}1\rangle\langle YABii{+}1\rangle\langle YAB1j\rangle\langle YAB1j{+}1\rangle\\\langle YABjj{+}1\rangle\langle YAB1k\rangle\langle YAB1k{+}1\rangle\langle YABkk{+}1\rangle\end{array}},
 \end{split}
\end{equation}
and
\begin{equation}
 \begin{split}
[1ii+1]&=\frac{\begin{vmatrix} 
 \langle YAn12\rangle&\langle YAi{-}1ii{+}1\rangle\langle YAii{+}1i{+}2\rangle
\\ \langle ABn12\rangle&\langle ABi{-}1ii{+}1\rangle\langle ABii{+}1i{+}2\rangle\\ \langle BYn12\rangle&\langle BYi{-}1ii{+}1\rangle\langle BYii{+}1i{+}2\rangle
\end{vmatrix}^2}{\begin{array}{c}\langle YAB(n12){\cap}(i{-}1ii{+}1)\rangle\langle YAB(i{-}1ii{+}1){\cap}(ii{+}1i{+}2)\rangle\\\langle YAB(ii{+}1i{+}2){\cap}(n12)\rangle\end{array}}.
 \end{split}
\end{equation}
We have checked numerically that this representation matches the corresponding BCFW representation of the integrand.
\subsection{All multiplicity generalization}
To go to the higher multiplicity case, we need to further triangulate each sign flip cell relevant for the $\mathcal{Y}$ amplituhedron. Let us consider, for example, the $\mathcal{A}_{234}$ cell for $n=7$, where the labelling means that the cell has three sign flips at $\langle \mathcal{Y} 12 \rangle,\langle \mathcal{Y}13 \rangle$ and $\langle \mathcal{Y} 14 \rangle$. To obtain the vertices of the intersecting polygon, we need to triangulate by considering the signs of additional brackets such as $\langle \mathcal{Y} ij \rangle$ where $j\neq i{+}1$ and $i\neq1$. In this case, there are ten possible sign patterns and therefore a priori up to ten distinct polygons for each cell. Na\"{i}vely, the increase in both the number of cells and the complexity of the intersecting polygons for each cell makes identifying all-multiplicity structure difficult. However, we have already seen that the 6$\times$2 representation of the six-point space is relatively simple (\ref{eq:6pt62rep}) -- in fact, this structure persists to the higher multiplicity cases! By explicit calculation similar to that of the previous section and appendix~\ref{sec:appendix1}, we find the seven point amplituhedron can be triangulated as the sum of the following spaces:
\begin{align}
\label{eq:7ptform}
\mathcal{A}^{\text{1-loop}}_{7,1}=&\mathcal{A}^{m=2}_{7,3}(1,\ldots,7)\times \mathcal{A}^{m=2}_{7,1}(\hat{1},\ldots,\hat{7})\nonumber\\
&+\sum_{\substack{1\leq i\leq 7 \\ 2\leq k\leq 3}}\mathcal{A}_{8-k,3}^{m=2}(i+k,i+k+1,\ldots,i)\nonumber\\
&\times\Big[\mathcal{A}^{(2)}((ii{+}ki{-}1),(ii{+}ki{+}k{-}1),(ii{+}ki{+}1))\nonumber\\&\quad\quad{+}\mathcal{A}^{(2)}((ii{+}ki{-}1),(ii{+}ki{+}1),(ii{+}ki{+}k{+}1))\Big].
 \end{align}
Although we will not derive this result (or its higher multiplicity generalizations), it can be easily checked by simply computing the canonical form (which we shall write below for arbitrary multiplicities) and comparing to the literature. For the eight point case we find a very similar expression, although the $k=3$ amplituhedron involves a slightly more complicated triangulation which is representative of the $n$-point case: 
\begin{align}
\label{eq:8ptform}
\mathcal{A}^{\text{1-loop}}_{8,1}=&\mathcal{A}^{m=2}_{8,3}(1,\ldots,8)\times \mathcal{A}^{m=2}_{8,1}(\hat{1},\ldots,\hat{8})\nonumber\\
&+\sum_{\substack{1\leq i\leq 8\\ 2\leq k\leq 4}}\left[\mathcal{A}_{(9{-}k),3}^{m=2}(i{+}k,i{+}k{+}1,\ldots,i){+}\mathcal{A}_{(k{+}1),3}^{m=2}(i,i+1,\ldots,i{+}k)\right]\nonumber\\
&\times\Big[\mathcal{A}^{(2)}((ii{+}ki{-}1),(ii{+}ki{+}k{-}1),(ii{+}ki{+}1))\nonumber\\&\quad\quad+\mathcal{A}^{(2)}((ii{+}ki{-}1),(ii{+}ki{+}1),(ii{+}ki{+}k{+}1))\Big].
 \end{align}
From these results, the natural conjecture for the 6$\times$2 representation of the $n$-point amplituhedron space is
\begin{align}
\label{eq:nptspace}
\mathcal{A}^{\text{1-loop}}_{n,1}=&\mathcal{A}^{m=2}_{n,3}(1,\ldots,n)\times \mathcal{A}^{m=2}_{n,1}(\hat{1},\ldots,\hat{n})\nonumber\\
&+\frac{1}{2}\sum_{\substack{1\leq i\leq n \\ 2\leq k\leq n{-}2}}\left[\mathcal{A}_{n{-}k{+}1,3}^{m=2}(i{+}k,i{+}k{+}1,\ldots,i){+}\mathcal{A}_{k{+}1,3}^{m=2}(i,i{+}1,\ldots,i{+}k)\right]\nonumber\\
&\times\Big[\mathcal{A}^{m=2}_{3,1}((ii{+}ki{-}1),(ii{+}ki{+}k{-}1),(ii{+}ki{+}1))\nonumber\\&\quad\quad+\mathcal{A}^{m=2}_{3,1}((ii{+}ki{-}1),(ii{+}ki{+}1),(ii{+}ki{+}k{+}1))\Big].
 \end{align}
 The associated canonical form can now be written down trivially using the $m=2$ result (\ref{eq:m3nkForm}) for the all-multiplicity, all-helicity amplituhedron canonical form. This gives our result, the $6\times2$ representation for the NMHV one-loop integrand, expressed as the product of a form in the plane $\mathcal{Y}$ and the point $y$ on the intersecting polygon:
 \begin{align}
\label{eq:nptform}
\Omega^{\text{1-loop}}_{n,1}=&\Omega^{m=2}_{n,3}(1,\ldots,n)\times \Omega^{m=2}_{n,1}(\hat{1},\ldots,\hat{n})\nonumber\\
&+\frac{1}{2}\sum_{\substack{1\leq i\leq n \\ 2\leq k\leq n{-}2}}\left[\Omega_{n{-}k{+}1,3}^{m=2}(i{+}k,i{+}k{+}1,\ldots,i){-}\Omega_{k{+}1,3}^{m=2}(i,i{+}1,\ldots,i{+}k)\right]\nonumber\\
&\times\Big[\Omega^{(2)}((ii{+}ki{-}1),(ii{+}ki{+}k{-}1),(ii{+}ki{+}1))\nonumber\\&\quad\quad+\Omega^{(2)}((ii{+}ki{-}1),(ii{+}ki{+}1),(ii{+}ki{+}k{+}1))\Big],
 \end{align}
where $\Omega^{(2)}$ is the two-form for a triangle. We have checked that this formula is consistent with the BCFW result  \cite{ArkaniHamed:2010kv} up to (and including) twenty-two points numerically. This canonical form is expressed as a product of a six-form for the $k=3$ space and a two-form for the intersecting polygon, which corresponds to $k=1$. This is to be contrasted with all other known representations of the integrand, which are written as a product of $R$-invariants and MHV one-loop integrands. 

Our result (\ref{eq:nptform}) has term-by-term spurious poles which are associated to the spurious lines $\hat{i}\hat{j}$ used to triangulate the intersecting polygon.  However, triangulating the polygon by introducing spurious points such as $(\hat{i}\hat{i}{+}1){\cap}(\hat{j}\hat{j}{+}1)$ would instead lead to a $6{\times}2$ representation which, when written back in $(Y,(YAB))$ space, has only $\langle Yii{+}1jj{+}1\rangle$ and $\langle YABii{+}1\rangle$ poles. By construction, such a representation would be ``super-local'' in the sense of \cite{Arkani-Hamed:2014dca}, and might have interesting positivity properties relevant for the yet-to-be-understood dual of the amplituhedron. We leave a detailed investigation of this topic for future work.

\section{Conclusion}
\label{sec:5}
In this work, we have begun the systematic investigation of the all-multiplicity $m=4$ tree and loop-level amplituhedron for the next-to and next-to-next-to maximally helicity violating configurations. The topological characterization of the amplituhedron replaces the computation of scattering amplitudes and loop integrands in planar $\mathcal{N}=4$ sYM by a simple to state (but extremely nontrivial) geometry problem. For the NMHV and $\text{N}^2$MHV tree-level cases, the natural triangulation associated to the sign-flip definition is not directly related to the BCFW recursion or any other previously known representation of the amplitude. However, the canonical forms associated to individual sign flip patterns are significant, and moreover there seems to be some rough correspondence between different helicity sectors. In addition to pushing to higher $k$ at tree-level, it would be interesting to make the connection between the different helicity sectors more precise, as was done for the $m=2$ geometry. Another avenue for future exploration is in the classification of different sign flip patterns. At the $\text{N}^2$MHV level, we found that some sign flip patterns had more complicated geometries (and associated forms) than others; a general understanding of the correspondence between the inequalities needed to define the space and the actual boundary structure in the canonical form would likely lead to significant progress in the triangulation problem.

At loop-level, we constructed the $6{\times}2$ representation of the one-loop NMHV amplituhedron from the sign flip characterization of the space. This representation is an immediate consequence of the topological definition and realizes the one-loop space as the intersection of the $m=4$ tree-level and $m=2,k=3$ geometries. The triangulation suggested by the $6{\times}2$ picture is a dramatic departure from the usual way of thinking about the NMHV one-loop integrand. In the future it would be interesting to examine the structure of our result in the original momentum twistor space. More generally, the $2(k{+}2)\times2k$ representation of the one-loop $\text{N}^k$MHV geometry seems to offer a clear path forward to extending the results of this paper. Although for higher $k$ the geometry is much richer, the one-loop space can always be constructed from two $m=2$ amplituhedra, where the triangulation problem is under significantly more control, suggesting the problem might be solvable for arbitrary $n$ and $k$. Another future direction is to go to higher loops. For example, the sign flip characterization of the two-loop NMHV amplituhedron suggests the relevant spaces are $m=2,k=3$ amplituhedra in two planes $(YAB),(YCD)$ and the two associated intersecting polygons -- where we now have the additional mutual positivity condition $\langle YABCD\rangle>0$ between loops.

\section*{Acknowledgements}
We thank Jaroslav Trnka for first suggesting the problem and providing guidance and encouragement throughout the project. We also thank Song He for numerous stimulating discussions. This work was supported through the hospitality of the Center for Quantum Mathematics and Physics (QMAP), Department of Physics, University of California, Davis and the Institute of Theoretical Physics, Chinese Academy of Sciences (ITP-CAS). This work is supported by the SOKENDAI Long-term Internship Program. The research of CL is supported in part by U.S. Department of Energy grant DE-SC0009999 and by the funds of the University of California.

\appendix
\section{6$\times$2 Representation of the six-point integrand}
\label{sec:appendix1}
For the six-point case, there are four sign flip cells $\mathcal{A}_{234},\mathcal{A}_{235},\mathcal{A}_{245},\mathcal{A}_{345}$. In section~\ref{sec:6ptNMHV1loop} we triangulated the $\mathcal{A}_{234}$ cell; in this appendix, we complete the six-point calculation. The vertices of the polygon which intersect with the $\mathcal{A}_{235}$ cell are
\begin{equation}
(123), (234), (561),
\end{equation}
while the other vertices (which depend on the signs of additional brackets) are given by the list of cases
\begin{center}
\begin{tabular}{| c | c | c | c | c | c | c |}
\hline
 $(25)$&$(35)$&$(26)$&$(36)$&$(46)$&vertices&pentagon\\\hline
+&$-$&+&+ & $-$ & &\\\cline{1-5}
$-$&$-$&+&+ & $-$ &\multirow{2}{*}{$(345),(456),(612)$}&\multirow{2}{*}{$(4)$}\\\cline{1-5}
+&$-$&+&$-$ & $-$ &&\\\cline{1-5}
$-$&$-$&+&$-$ & $-$ &&\\\cline{1-7}
+&$-$&+&$-$ & +&\multirow{2}{*}{$(345),(346),(612),(461)$}&\multirow{2}{*}{$(5)$}\\\cline{1-5}
$-$&$-$&+&$-$ & +&&\\\cline{1-7}
$-$&+&+&$-$ & + &$(235),(346),(612),(356),(461)$&$(6)$\\\cline{1-7}
+&$-$&$-$&+ & $-$ &$(236),(345),(456),(256)$&$(7)$\\\cline{1-7}
$-$&+&+&$-$ & $-$ &$(235),(456),(612),(356)$&$(8)$\\\cline{1-7}
\end{tabular}
\end{center}
\begin{figure}[t]
\begin{center}
\includegraphics[clip,width=14.0cm]{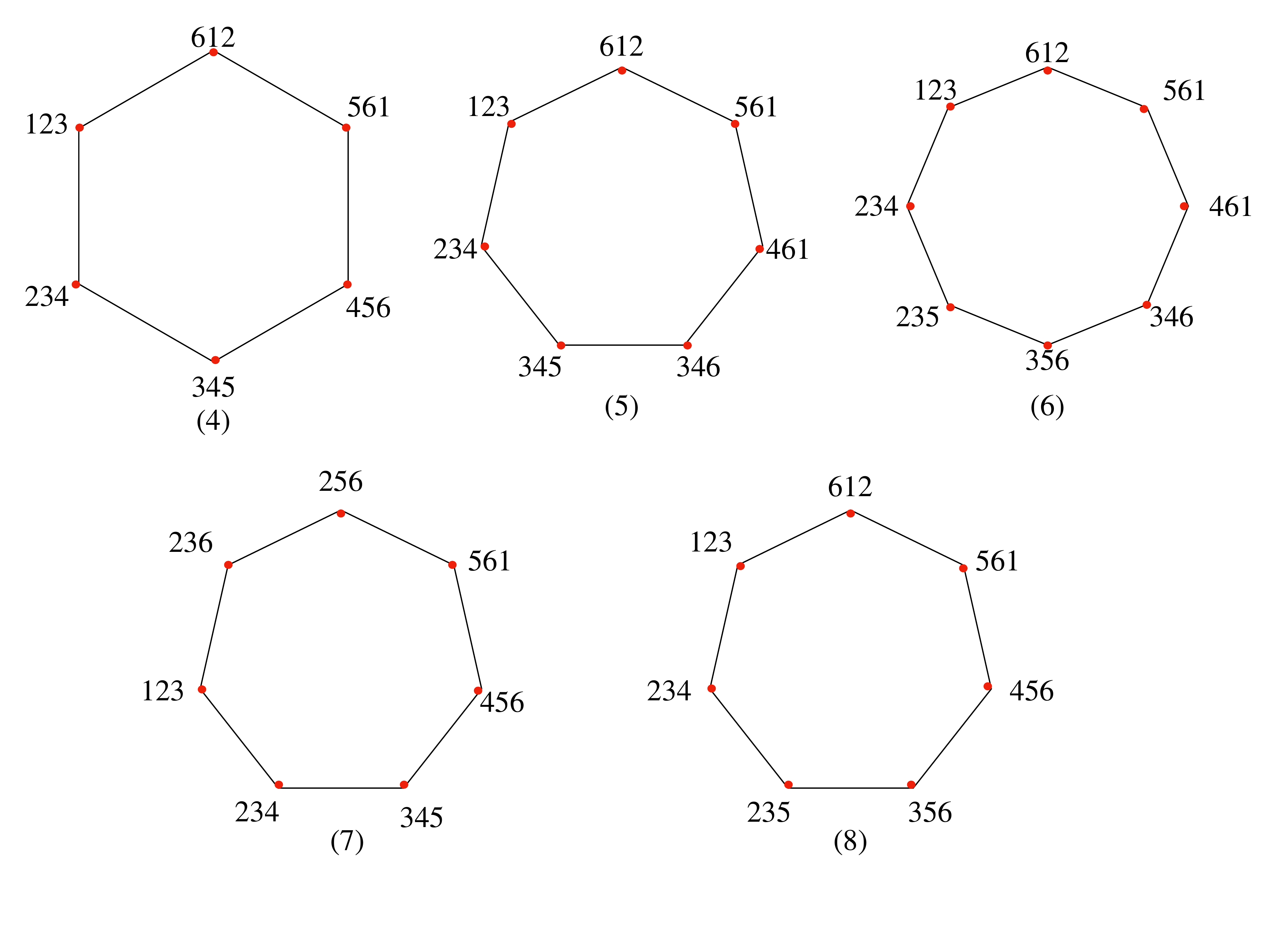}
\end{center}
\caption{Polygons for $\mathcal{A}_{235}$}
 \label{fig:pentagon235cell}
\end{figure}
where we use the shorthand notation $(ij)=\langle Yij\rangle$. For each case, we draw the associated intersecting polygon in Figure~\ref{fig:pentagon235cell}. Next, the vertices of the polygon which intersects with $\mathcal{A}_{245}$ cell are
\begin{equation}
(123),(456),(561)
\end{equation}
and for the additional vertices we have the different cases
\begin{center}
\begin{tabular}{| c | c | c | c | c | c | c |}
\hline
 $(24)$&$(25)$&$(35)$&$(26)$&$(36)$&vertices&pentagon\\\hline
+&+&$-$&+ & + & &\\\cline{1-5}
$-$&$-$&$-$&+ & + &\multirow{2}{*}{$(234),(345),(612)$}&\multirow{2}{*}{$(9)$}\\\cline{1-5}
$-$&+&$-$&+ & $-$ &&\\\cline{1-5}
$-$&$-$&$-$&+ & $-$ &&\\\cline{1-7}
+&$-$&$-$&+ & +&\multirow{2}{*}{$(124),(345),(245),(612)$}&\multirow{2}{*}{$(10)$}\\\cline{1-5}
+&$-$&$-$&+ & $-$&&\\\cline{1-7}
+&$-$&+&+ & $-$ &$(124),(235),(245),(612),(356)$&$(11)$\\\cline{1-7}
$-$&$-$&+&+ & $-$ &$(234),(235),(612),(356)$&$(12)$\\\cline{1-7}
$-$&+&$-$&$-$ & + &$(234),(236),(345),(256)$&$(13)$\\\cline{1-7}
\end{tabular}
\end{center}
\begin{figure}[t]
\begin{center}
\includegraphics[clip,width=14.0cm]{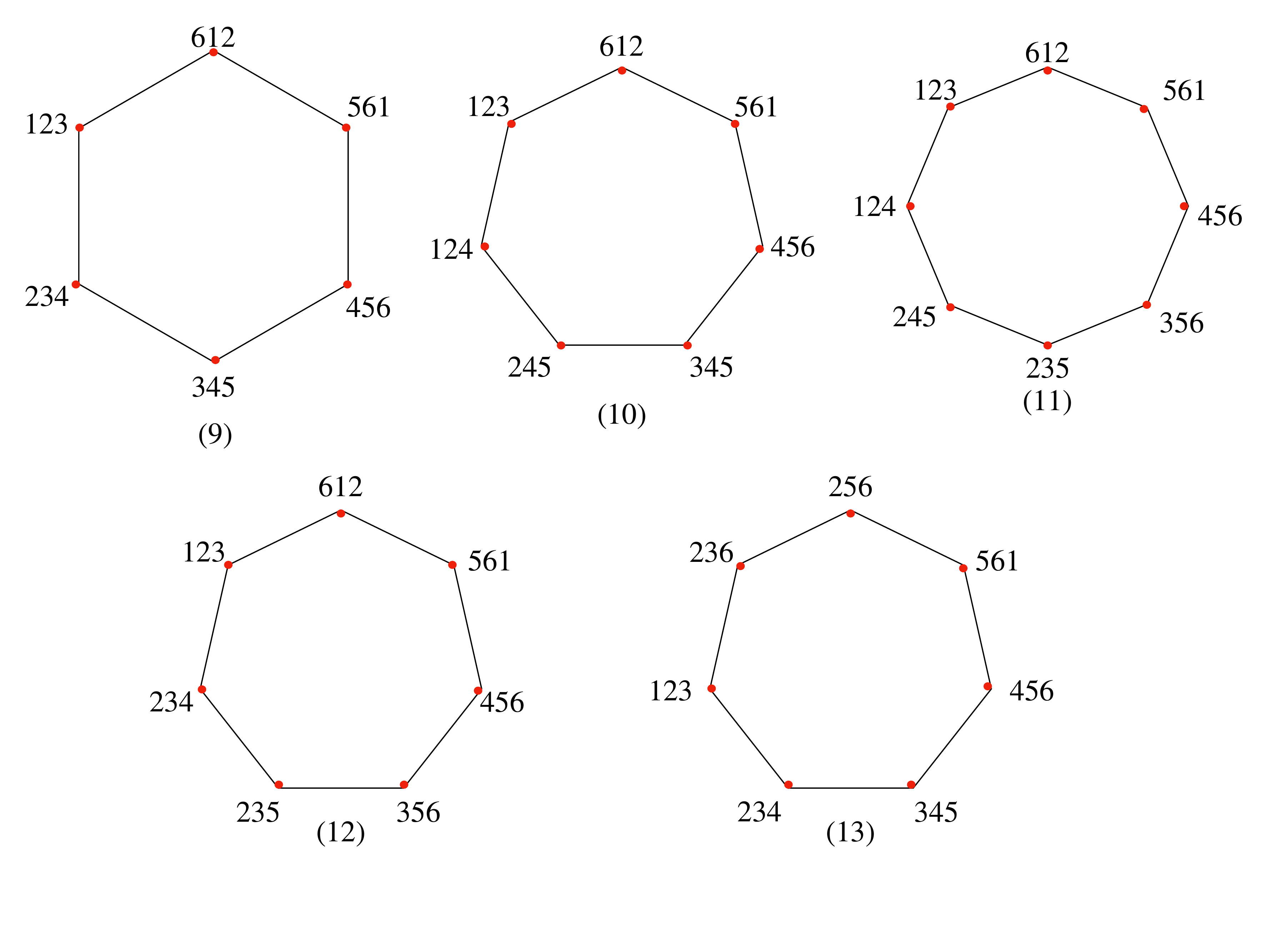}
\end{center}
\caption{Polygons for $\mathcal{A}_{245}$}
 \label{fig:pentagon245cell}
\end{figure}
and the associated intersecting polygons are given in Figure~\ref{fig:pentagon245cell}.
Finally, the vertices of the polygon which intersects with $\mathcal{A}_{345}$ cell are
 \begin{equation}
(134), (345),(456), (561),(136)
\end{equation}
and the additional vertices are given in the table
\begin{center}
\begin{tabular}{| c | c | c | c | c |}
\hline
 $(24)$&$(25)$&$(26)$&vertices&pentagon\\\hline
$-$&+&+&\multirow{2}{*}{$(234),(612)$}&\multirow{2}{*}{$(14)$}\\\cline{1-3}
$-$&$-$&+&&\\\cline{1-5}
+&$-$&+&$(124),(612),(245)$&$(15)$\\\cline{1-5}
$-$&+&$-$&$(234),(236),(256)$&$(16)$\\\cline{1-5}
\end{tabular}
\end{center}
\begin{figure}[t]
\begin{center}
\includegraphics[clip,width=14.0cm]{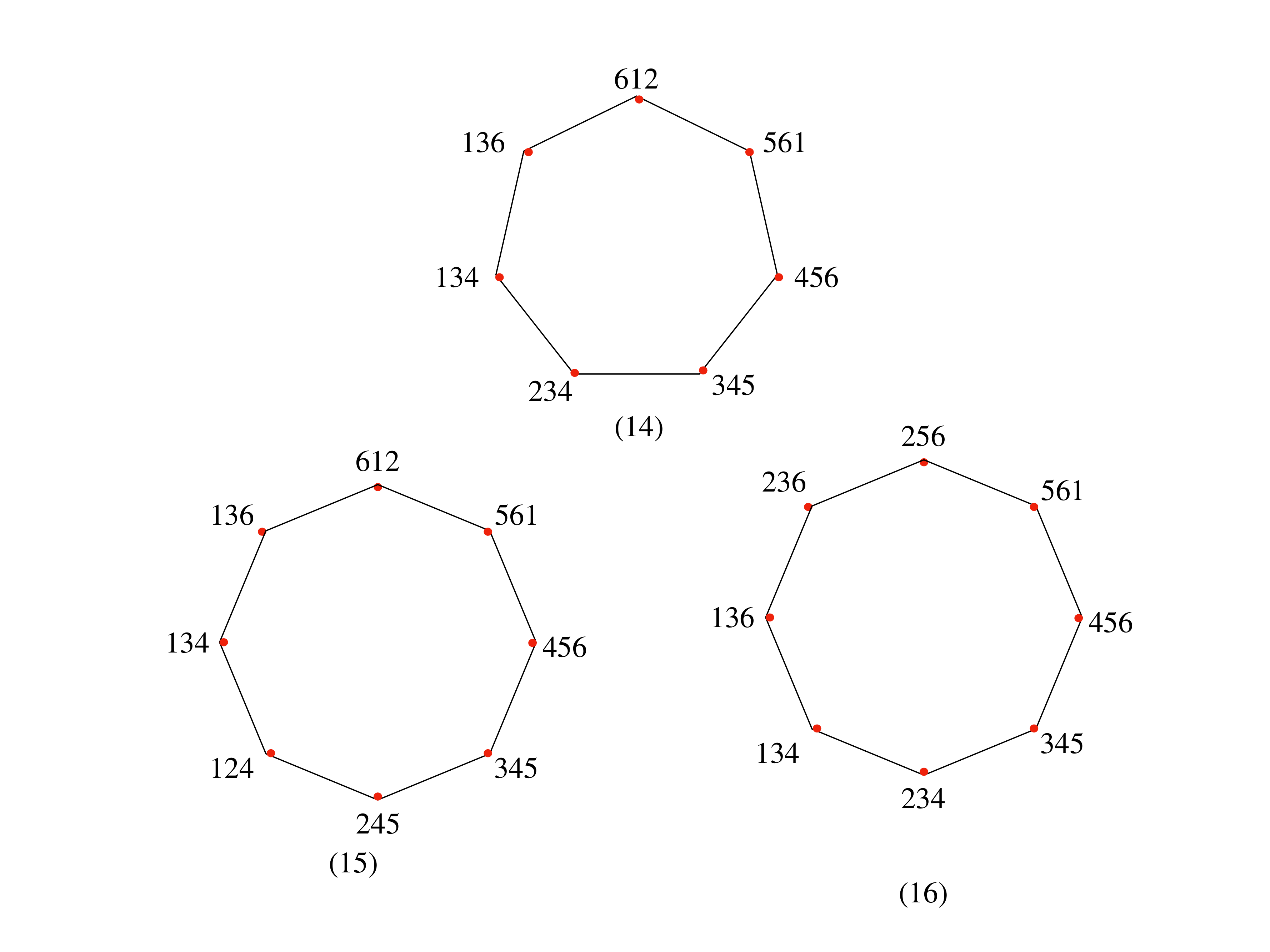}
\end{center}
\caption{Polygons for $\mathcal{A}_{345}$}
 \label{fig:pentagon345cell}
\end{figure}
and the associated intersecting polygons are given in Figure~\ref{fig:pentagon345cell}. Next, we consider the 6${\times}$2 triangulation of these subspaces. First, we can see that all the polygons labelled in Figures~\ref{fig:234polygon},\ref{fig:pentagon235cell},\ref{fig:pentagon245cell} and \ref{fig:pentagon345cell} are related to the basic polygon $P_6$ which has the six vertices $(612),(123),(234),(345),(456),(561)$ by the transformations
\begin{align}
&(1)=P_6-\Delta(561)(125)(145),\ \ \ (2)=P_6-\Delta(561)(125)(145)-\Delta(612)(236)(256),\nonumber\\
&(3)=P_6-\Delta(561)(125)(145)-\Delta(456)(461)(346),\ \ \ (5)=P_6-\Delta(456)(461)(346),\nonumber\\
&(6)=P_6-\Delta(345)(235)(356)-\Delta(456)(461)(346),\ \ \ (7)=P_6-\Delta(612)(236)(256),\nonumber\\
&(8)=P_6-\Delta(345)(235)(356),\ \ \ (9)=P_6-\Delta(234)(124)(245),\nonumber\\
&(11)=P_6-\Delta(234)(124)(245)-\Delta(345)(235)(356),\ \ \ (12)=P_6-\Delta(345)(235)(356),\nonumber\\
&(13)=P_6-\Delta(612)(236)(256),\ \ \ (14)=P_6-\Delta(123)(136)(134),\nonumber\\
&(15)=P_6-\Delta(123)(136)(134)-\Delta(234)(124)(245),\nonumber \\ &(16)=P_6-\Delta(123)(136)(134)-\Delta(612)(236)(256).\nonumber\\
 \end{align}
where $(i)$ is the pentagon $(i)$ and $\Delta(i)(j)(k)$ is the triangle whose vertices are $i,j,k$. From this, the 6$\times$2 representation of the six-point NMHV one-loop amplituhedron can be decomposed as
\begin{align}
\mathcal{A}^{6\times2}_{6,1}&=(\mathcal{A}_{234}+\mathcal{A}_{235}+\mathcal{A}_{245}+\mathcal{A}_{345})\times P_6+\mathcal{A}^1\times\Delta(612)(236)(256)\nonumber\\
&+\mathcal{A}^2\times\Delta(123)(136)(134)+\cdots+\mathcal{A}^6\times\Delta(561)(125)(145),
 \end{align}
where $\mathcal{A}^i$ is the union of the modified spaces $\mathcal{A}_{ijk}'$, defined as
\begin{align}
\mathcal{A}^{1}&=\mathcal{A}'_{234}+\mathcal{A}'_{235}+\mathcal{A}'_{245}+\mathcal{A}'_{345}\nonumber\\
\mathcal{A}'_{234}&:\mathcal{A}_{234}\ \  \text{with}\ \  \{(26),(36),(46)\}=\{-,+,-\}\nonumber\\
\mathcal{A}'_{235}&:\mathcal{A}_{235}\ \  \text{with}\ \  \{(25),(35),(26),(46)\}=\{+,-,-,-\}\nonumber\\
\mathcal{A}'_{245}&:\mathcal{A}_{245}\ \  \text{with}\ \  \{(25),(26),(36)\}=\{+,-,+\}\nonumber\\
\mathcal{A}'_{345}&:\mathcal{A}_{345}\ \  \text{with}\ \  \{(24),(25),(26)\}=\{-,+,-\}.
 \end{align}
Note that the signs of the brackets $\langle YABij \rangle$ defining this space are
\begin{equation}
\langle YABii+1 \rangle>0,\ \ \{\langle YAB62 \rangle,\langle YAB63 \rangle,\langle YAB64 \rangle,\langle YAB65 \rangle\}=\{+,-,+,-\},
\end{equation}
which is the sign flip condition of the five-point $m=2,k=3$ amplituhedron $\mathcal{A}^{m=2}_{5,3}(2,3,4,5,6)$. Similarly we can see that other subspaces $\mathcal{A}^i$ are simple relabelings of this space, namely:
\begin{align}
&\mathcal{A}^1=\mathcal{A}^{m=2}_{5,3}(2,3,4,5,6),\ \mathcal{A}^2=\mathcal{A}^{m=2}_{5,3}(3,4,5,6,1),\ \mathcal{A}^3=\mathcal{A}^{m=2}_{5,3}(4,5,6,1,2),\nonumber\\
&\mathcal{A}^4=\mathcal{A}^{m=2}_{5,3}(5,6,1,2,3),\ \mathcal{A}^5=\mathcal{A}^{m=2}_{5,3}(6,1,2,3,4),\ \mathcal{A}^6=\mathcal{A}^{m=2}_{5,3}(1,2,3,4,5).\nonumber\\
 \end{align}
From this, the final result of the 6$\times$2 representation of the canonical form \eqref{eq:6pt62rep} follows immediately. 
\bibliographystyle{jhep}
\bibliography{ref} 
\end{document}